%% file: article.tex
\documentclass[11pt,letterpaper]{article}

\usepackage[utf8]{inputenc}
\usepackage[letterpaper,margin=1in]{geometry}

\usepackage{caption}
\usepackage{subcaption}
\usepackage[inline]{enumitem}
\usepackage{float}
\usepackage{graphicx}
\usepackage{hyperref}
\usepackage{listings}
\usepackage{textcomp}
\usepackage{xcolor}
\usepackage{xurl}

\usepackage[backend=biber,style=numeric,sorting=none,giveninits=true,isbn=false,url=false,maxbibnames=99,citestyle=numeric-comp]{biblatex}
\renewbibmacro{in:}{}
\DeclareNameAlias{default}{family-given}

\usepackage[]{authblk}
\renewbibmacro{in:}{}

\definecolor{backcolour}{rgb}{0.9,0.91,0.91}
\definecolor{codegray}{rgb}{0.5,0.5,0.5}
\definecolor{codeblue}{rgb}{0,0,0.5}
\definecolor{codered}{rgb}{0.6,0,0}
\definecolor{codegreen}{rgb}{0,0.5,0}
\definecolor{codeyellow}{rgb}{0.5,0.5,0.0}

\lstdefinestyle{python}{%
    backgroundcolor=\color{backcolour},
    commentstyle=\color{codegray},
    keywordstyle=\color{codered},
    numberstyle=\tiny\color{codegray},
    stringstyle=\color{codegreen},
    basicstyle=\ttfamily\scriptsize,
    language=Python,
    breakatwhitespace=false,
    breaklines=true,
    captionpos=b,
    keepspaces=true,
    numbers=none,
    showspaces=false,
    showstringspaces=false,
    showtabs=false,
    tabsize=4,
    literate={->}{{\textcolor{codered}{->}}}{1}
}

\title{Highdicom: A Python library for standardized encoding of image annotations and machine learning model outputs in pathology and radiology}

\author[1,2]{Christopher P. Bridge}
\author[3]{Chris Gorman}
\author[4]{Steven Pieper}
\author[2]{Sean W. Doyle}
\author[5,6]{Jochen K. Lennerz}
\author[1,2,7]{Jayashree Kalpathy-Cramer}
\author[8]{David A. Clunie}
\author[7,9]{Andriy Y. Fedorov}
\author[3,6,*]{Markus D. Herrmann}

\affil[1]{%
    Martinos Center for Biomedical Imaging,
    Massachusetts General Hospital,
    Boston, MA, USA
}
\affil[2]{%
    MGH \& BWH Center for Clinical Data Science,
    Mass General Brigham,
    Boston, MA, USA
}
\affil[3]{%
    Computational Pathology,
    Department of Pathology,
    Massachusetts General Hospital,
    Boston, MA, USA
}
\affil[4]{%
    Isomics, Inc.,
    Cambridge, MA, USA
}
\affil[5]{%
    Center for Integrated Diagnostics,
    Department of Pathology,
    Massachusetts General Hospital,
    Boston, MA, USA
}
\affil[6]{%
    Department of Pathology,
    Harvard Medical School,
    Boston, MA, USA
}
\affil[7]{%
    Department of Radiology,
    Harvard Medical School,
    Boston, MA, USA
}
\affil[8]{%
    PixelMed Publishing, LLC.,
    Bangor, PA, USA
}
\affil[9]{%
    Surgical Planning Laboratory,
    Department of Radiology,
    Brigham and Women’s Hospital,
    Boston, MA, USA
}

\affil[*]{%
    corresponding author:
    \href{mailto:mdherrmann@mgh.harvard.edu}{mdherrmann@mgh.harvard.edu}
}

\date{\vspace{-5ex}}

\begin{document}

    \maketitle

    \begin{abstract}
        \noindent
        \input{./sections/abstract}
    \end{abstract}

    \section*{Background}
    \input{./sections/introduction}

    \section*{Methods}
    \input{./sections/methods}

    \section*{Results}
    \input{./sections/results}

    \section*{Discussion}
    \input{./sections/discussion}

    \section*{Conclusion}
    \input{./sections/conclusion}

    \section*{References}
    \printbibliography[heading=none]

\end{document}


\maketitle

\section*{Experimental image datasets}

\subsection*{Preparation of slide microscopy images and annotations in DICOM format}

We downloaded SM images in SVS format from either the Genomics Data Commons (GDC) Data Portal using the GDC Data Transfer Tool software or from TCIA via the CPTAC histopathology interface in case of TCGA and CPTAC collections, respectively.
In addition, we downloaded the corresponding biospecimen and clinical metadata in JSON format via the GDC Data Portal for both TCGA and CPTAC collections and additional metadata in JSON format via the CPTAC Clinical Data API for the CPTAC collections.
For each image contained in a given SVS file, we created a DICOM VL Whole Slide Microscopy Image instance and stored it in a DICOM Part 10 file.
Briefly, we copied image pixel data contained in TIFF tiles as well as relevant pixel-related metadata contained in TIFF tags (e.g., \texttt{ImageLength}, \texttt{ImageWidth}, \texttt{ImageDescription}, \texttt{Compression}, and \texttt{PhotometricInterpretation}) into the corresponding DICOM data elements.
We further enriched DICOM data sets with information extracted from TCIA biospecimen metadata JSON documents, e.g.\ fixatives and embedding media used for specimen preparation as well as relevant patient, study, and specimen identifiers.
For identifiers, we followed the same patterns used for the radiology data sets to facilitate matching of pathology and radiology DICOM data sets.

As described in the Methods section, we structured the content of the SR documents based on template TID 1500 ``Measurement Report'' and encoded image annotations using content items defined either in template TID 1410 ``Planar ROI Measurements and Qualitative Evaluations'' in case of graphical ROI annotations or TID 1501 ``Measurement and Qualitative Evaluation Group'' in case the annotations applied to an entire image or series.
In either case, we included content item ``Finding'' (DCM 121071) to encode the concept that the given image or image region represents, e.g., ``Neoplasm'' (SCT 108369006).
We further extended template TID 1501 and included additional content items to encode measurements and qualitative evaluations of the image or image region that are specific to cancer diagnosis in pathology or radiology.
We included content items ``Morphology'' (SCT 116676008) and ``Topography'' (SCT 116677004) with value type \texttt{CODE} to encode the tumor histomorphology (squamous cell carcinoma, adenocarcinoma, etc.) and tissue of origin (bronchus, upper lobe, etc.) using the International Classification of Diseases for Oncology (ICD-O-3) and the Clinical Modification of the International Classification of Diseases (ICD-10-CM) coding systems, respectively.
In addition, we included content items ``Percent tumor cells'' (caDSR 5432686), ``Percent tumor nuclei'' (caDSR 5455534), and ``Specimen necrosis'' (caDSR 5455511) with value type \texttt{NUM} to provide measurements of the percentage of viable or dead tumor tissue.
We encoded graphical annotations of image regions of interest (ROIs) as vector graphics in DICOM Comprehensive SR or Comprehensive 3D SR documents as described above.
For SM images used in the pathology experiments, graphical annotations of tumor regions were not available in TCIA, and we therefore encoded image-level annotations in DICOM Comprehensive SR documents using TID 1501 ``Measurement and Qualitative Evaluation Group''.
However, we annotated a subset of SM images using the SliM viewer~\footnote{\url{https://github.com/mghcomputationalpathology/slim}} and encoded the resulting ROIs as content items with type \texttt{SCOORD3D} in DICOM Comprehensive 3D SR documents using TID 1410 ``Planar ROI Measurements and Qualitative Evaluations''.

\subsection*{Preparation of computed tomography images and annotations in DICOM format}

For the radiology experiments, we used the LIDC-IDRI dataset of 1018 diagnostic and screening CT studies from multiple institutions.
We downloaded CT images, consisting of a single axial series per study, in DICOM format from TCIA using the NBIA Data Retriever software.
Each scan in the LIDC-IDRI dataset is annotated in the form of segmentation masks for lung nodules.
Each scan contains one or more annotated nodules, and each nodule may be annotated by multiple readers.
The annotations for this dataset were already available in the original custom XML-based format, and also as DICOM Segmentation images and DICOM SR documents that were created in prior work using the C++-based DCMQI toolkit~\cite{Fedorov2018,Fedorov2020}.
For the purpose of demonstrating a full workflow within the Python programming language, we re-created annotations in both DICOM Segmentation and SR format from the original XML annotations, largely following Fedorov et al.~\cite{Fedorov2018,Fedorov2020}, using \textit{highdicom} and used the resulting datasets (rather than those created with DCMQI) for the subsequent experiments.
The SEG images are hereby used to encode the annotation of a single nodule by a single reader as a single segment, with a Segmented Property Category of ``Morphologically Abnormal Structure'' (SCT 49755003) and Segmented Property Type ``Nodule'' (SCT 27925004).
The SR documents used the Comprehensive3DSR IOD and included measurements and qualitative evaluations of each of the regions in the SEG images, referenced via a content item of type \texttt{IMAGE}.
Measurements of each nodule ROI included the ``Volume'' (SCT 118565006) and ``Diameter'' (SCT 81827009) as \texttt{NUM} content items, and qualitative evaluations included the subtlety, internal structure, calcification, sphericity, margin, lobulation, spiculation, texture, and malignancy of the nodule using a range of coded concepts as suggested by~\cite{Fedorov2018,Fedorov2020}.
Note that these measurements and qualitative evaluations were not used during model training, but were included in the annotation SR for the sake of completeness.
The script used for conversion is publicly available\footnote{\url{https://github.com/QIICR/lidc2dicom}}.

\subsection*{Validation of DICOM data sets}

We evaluated the compliance of preparated data sets with the DICOM standard using both automated validation tools and manual expert review.
Specifically, we used the \texttt{dciodvfy} command line tool of the \emph{dicom3tools}~\footnote{\url{http://www.dclunie.com/dicom3tools.html}} package to assert that individual DICOM files are structured according the corresponding Information Object Definition (IOD) and the \texttt{DicomSRValidator} program of the \emph{PixelMed Java DICOM Toolkit}~\footnote{\url{http://www.pixelmed.com/dicomtoolkit.html}}.
We further used the \texttt{dcdump} and \texttt{dcsrdump} command line tools of the \emph{dicom3tools} package and the \texttt{dcmdump} command line tool of the \emph{DICOM toolkit (DCMTK)}~\footnote{\url{https://dcmtk.org/}} package to manually review and validate the content of DICOM files.

\section*{Further model training details}

We implemented the deep convolutional neural network models using the \emph{PyTorch} Python library~\cite{NIPS2019.9015} and trained them on a Linux supercomputer with NVIDIA V-100 graphical processing units (GPUs) using the \emph{CUDA} and \emph{cuDNN} C++ libraries.
We optimized model parameters using Adam optimizer with momentum and optimized the learning rate as well as other hyperparameters using random search.

The data preprocessing pipelines were implemented in form of classes derived from \texttt{torch.data.Dataset}.
 which load the images and corresponding image annotations from DICOM files into \emph{NumPy} arrays, transform the data in memory into the representation expected by the respective model, and return a pair of image frames and labels as \emph{PyTorch} tensors.
Instances of these classes are instantiated given the location of DICOM files on disk as well as the SOP Instance UIDs of DICOM SR documents, which contain annotations that are considered the ground truth for model training as well as references to the source images from which the annotations were derived.
For each example, the \texttt{\_\_getitem\_\_} method loads image frames and annotations from DICOM files on disk into memory as \texttt{numpy.ndarray} objects, transforms the data into the representation expected by the first layer of the neural network (optionally performing data augmentation), and returns the pair of transformed pixel data and associated labels as a tuple of \texttt{torch.tensor} objects.

\section*{Model evaluation}

We selected one pathology and radiology model on a validation set and evaluated the performance of selected models on a hold-out test set.

\begin{figure}[h]
    \includegraphics[width=\textwidth]{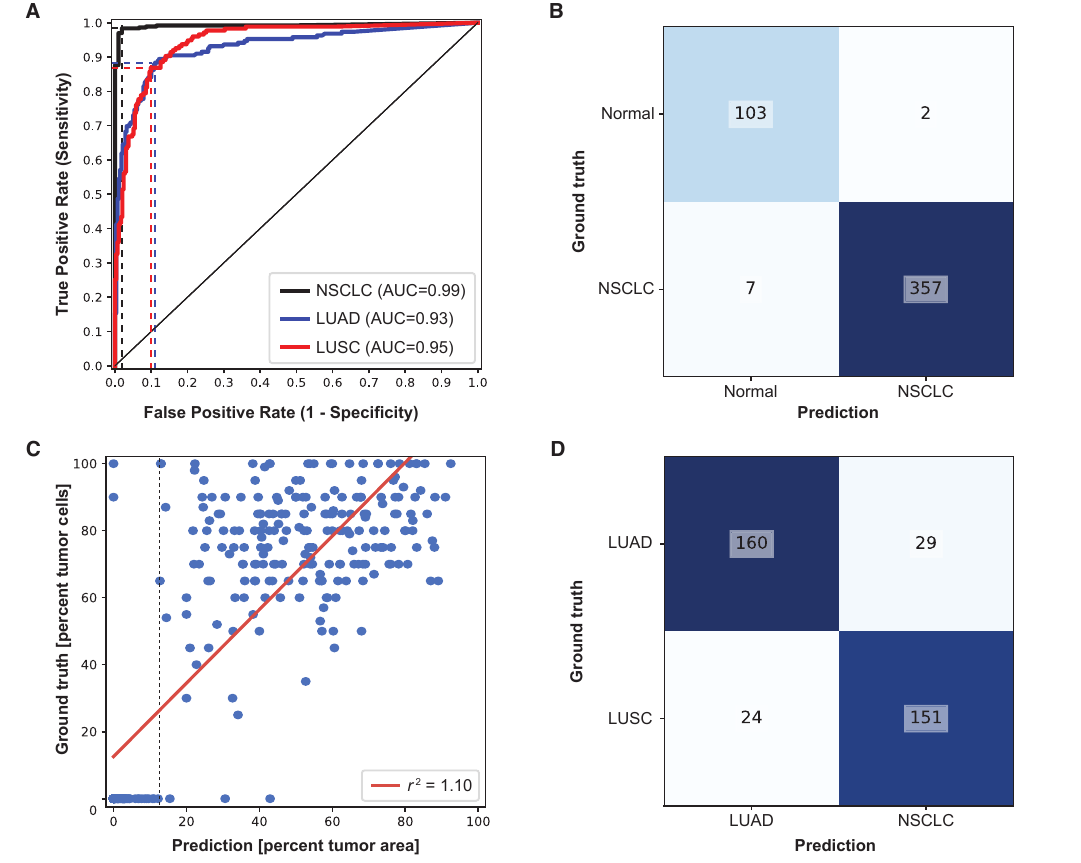}
    \caption{%
        Performance of SM image classifiers (NSCLC --- non-small cell lung cancer, LUAD --- lung adenocarcinoma, LUSC --- lung squamous cell carcinoma).
        \textbf{A} Receiver operating characteristic curve for each binary classification problem.
        \textbf{B} Confusion matrix comparing the ground truth against predicted class labels for the classification of normal lung versus non-small cell lung cancer using the threshold selected in A.
        \textbf{C} Correlation between annotated percent tumor cells and predicted percent tumor area considering examples that were classified as non-small cell lung cancer.
        \textbf{D} Confusion matrix comparing the ground truth against predicted class labels for the classification of lung adenocarcinoma versus lung squamous cell carcinoma considering examples that were correctly classified as non-small cell lung cancer.
    }\label{fig:sm-evaluation}
\end{figure}

For pathology, we developed classifiers to categorize whole slide images into either normal lung tissue, lung adenocarcinoma, or lung squamous cell carcinoma.
To this end, we thresholded the probabilistic fractional segmentation images outputted by the neural network model (using a threshold value of 0.5 probability) for each class and used thereby generated binary segmentation masks to compute the relative tumor area, i.e., the ratio of the number of image frames classified as tumor and the number of image frames classified as tissue.
Based on these aggregated measurements, we created binary whole slide image classifiers to distinguish normal lung from non-small cell lung cancer (NSCLC, lung adenocarcinoma or lung squamous cell carcinoma), lung adenocarcinoma (LUAD), or lung squamous cell carcinoma (LUSC).
The optiomal threshold for each classifier was determined via Receiver Operator Characteristic (ROC) analysis (Supplementary Figure~\ref{fig:sm-evaluation}A).
Evaluating the classifiers at the selected threshold values, we achieved an accuracy of 0.98 for separating normal lung from NSCLC (Supplementary Figure~\ref{fig:sm-evaluation}B), a high correlation between the predicted relative tumor area and the annotated tumor cell percentage for NSCLC examples (Supplementary Figure~\ref{fig:sm-evaluation}C), and an accuracy of 0.85 for distinguishing between LUAD and LUSC (Supplementary Figure~\ref{fig:sm-evaluation}D).

\begin{figure}[h]
    \includegraphics[width=\textwidth]{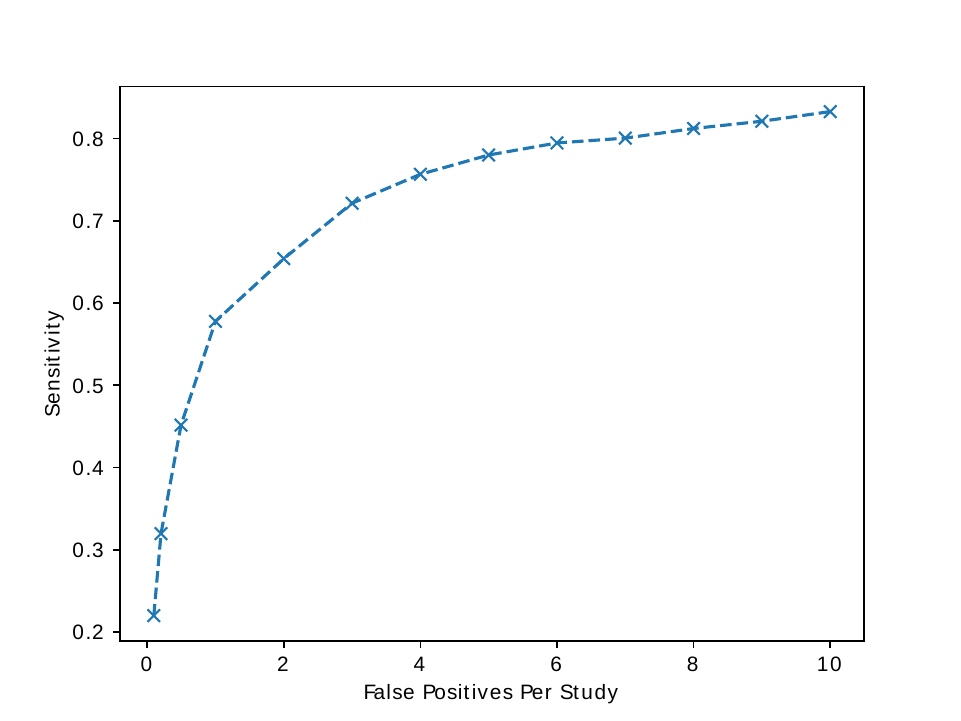}
    \caption{%
        Free-response receiver operating characteristic for the CT lung nodule detection model.
    }\label{fig:ct-evaluation}
\end{figure}

For radiology, we evaluated the CT lung nodule detection model in terms of study-level results.
Nodule sensitivity (recall) was calculated as the fraction of annotated nodules for which any predicted box overlapped any reader's annotated with intersection-over-union (IoU) of 0.5 or greater on any frame.
False positives in neighboring frames were clustered together if their IoU (in the x and y directions only) was greater than 0.5, and assigned the score of the highest score in the cluster.
This gave an average precision metric of 0.582 for nodule detection.
The free-response receiver operating characteristic (FROC), which plots nodule sensitivity (recall) against the average number of false positives per study as the detection score threshold is varied (Supplementary Figure~\ref{fig:ct-evaluation}).

\section*{Visualization of Outputs}

\begin{figure}
    \begin{subfigure}{0.5\textwidth}
        \centering
        \includegraphics[width=0.95\textwidth]{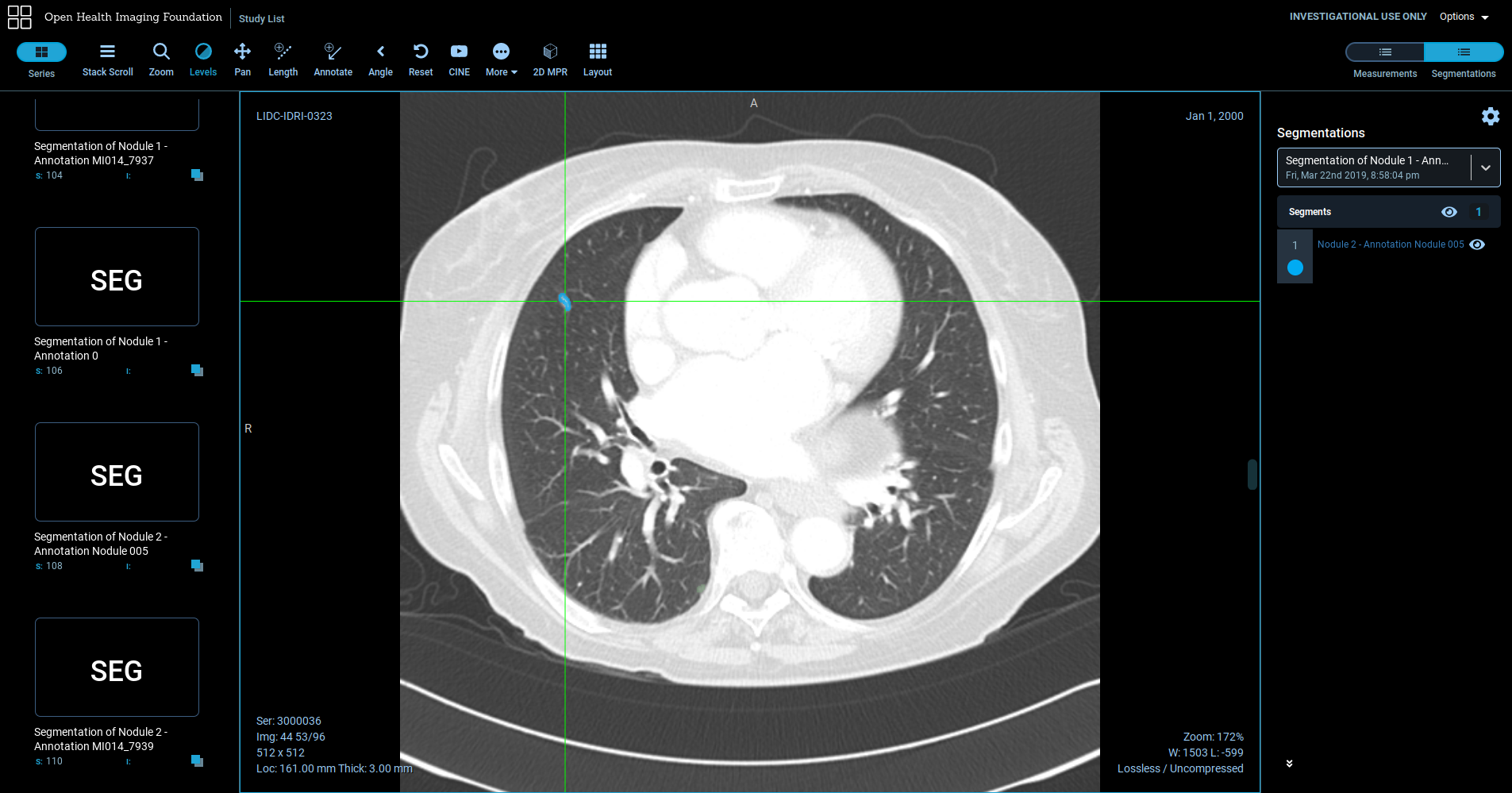}
    \end{subfigure}
    \begin{subfigure}{0.5\textwidth}
        \centering
        \includegraphics[width=0.95\textwidth]{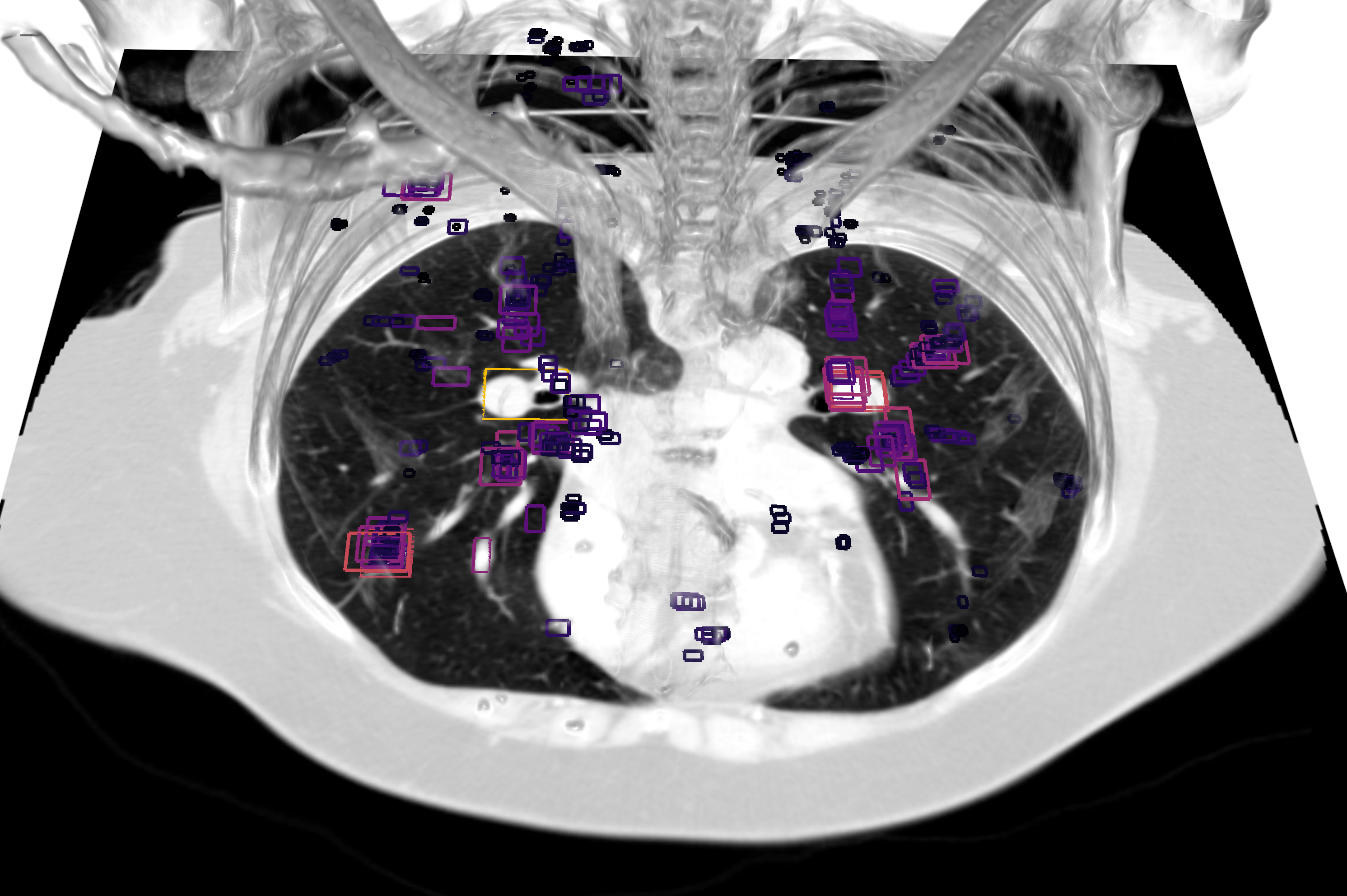}
    \end{subfigure}
    \caption{%
        \textbf{A} Lung nodule annotation in a CT image of the LIDC-IDRI collection encoded as a segment in a DICOM Segmentation image using highdicom and displayed in the open-source OHIF viewer.
        \textbf{B} Bounding box detection output from the CT lung nodule detection model encoded as a DICOM SR document and visualized using the open-source 3D Slicer software and the Quantitative Reporting extension~\cite{fedorov20121323}.
        Detected bounding boxes are shaded by their detection scores from purple (very low score) to yellow (high score).
    }\label{fig:ct-ohif-slicer}
\end{figure}

\begin{figure}
    \includegraphics[width=\textwidth]{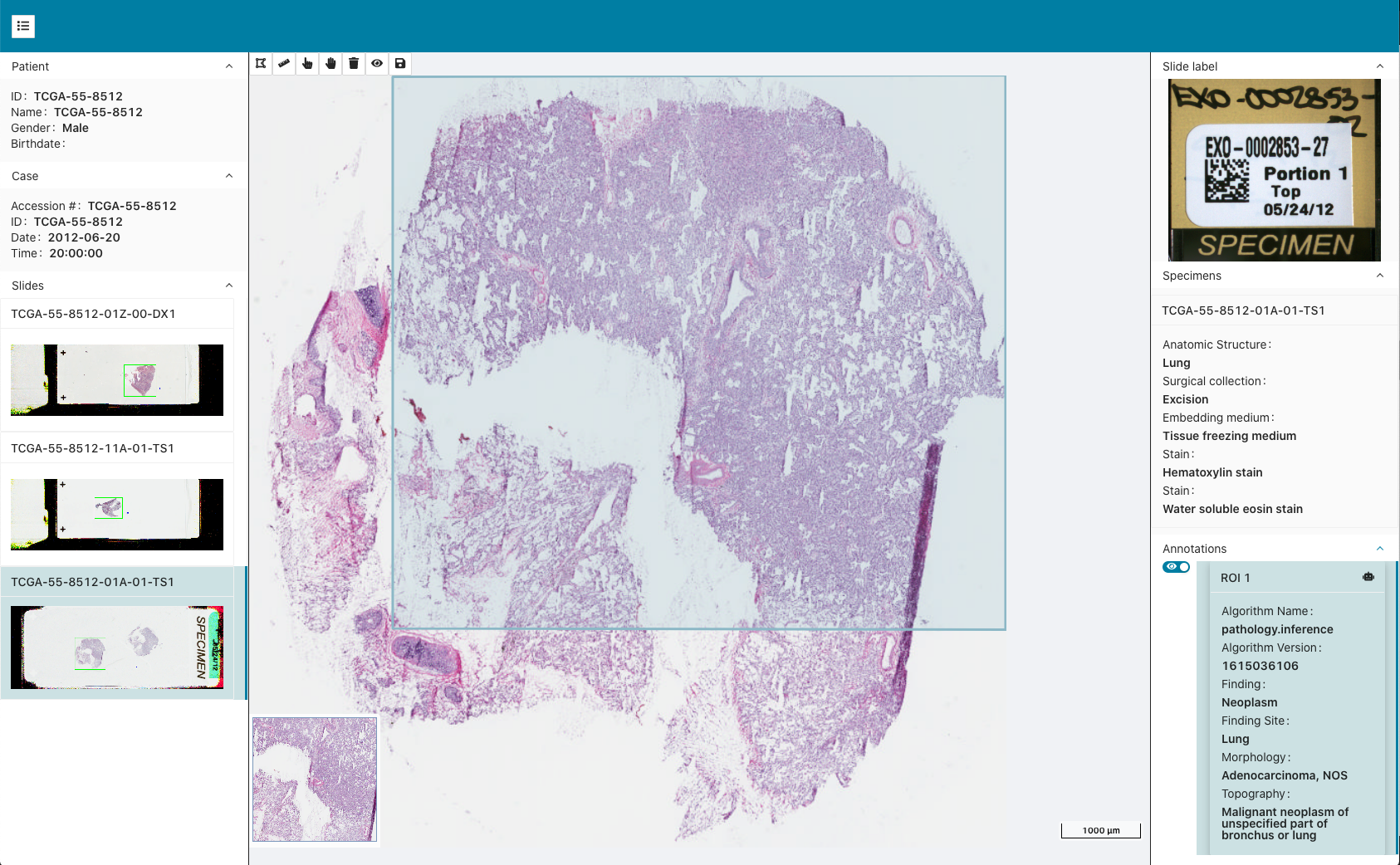}
	\caption{Lung adenocarcinoma regions detected in a slide microscopy image of the TCGA-LUAD collection encoded as SCOORD3D content items in a DICOM Comprehensive 3D SR document using \textit{highdicom} and displayed in the open-source SliM viewer.}\label{fig:sm-slim}
\end{figure}

Figures \ref{fig:ct-ohif-slicer} and \ref{fig:sm-slim} contain examples of image annotations and model outputs visualized within open source tools.

\section*{References}
\printbibliography[heading=none]

%% file: sections/abstract.tex
Machine learning (ML) is revolutionizing image-based diagnostics in pathology and radiology.
ML models have shown promising results in research settings, but the lack of interoperability between ML systems and enterprise medical imaging systems has been a major barrier for clinical integration and evaluation.
The DICOM\textsuperscript{\textregistered} standard specifies Information Object Definitions (IODs) and Services for the representation and communication of digital images and related information, including image-derived annotations and analysis results.
However, the complexity of the standard represents an obstacle for its adoption in the ML community and creates a need for software libraries and tools that simplify working with data sets in DICOM format.

Here we present the \emph{highdicom} library, which provides a high-level application programming interface (API) for the Python programming language that abstracts low-level details of the standard and enables encoding and decoding of image-derived information in DICOM format in a few lines of Python code.
The \emph{highdicom} library leverages NumPy arrays for efficient data representation and ties into the extensive Python ecosystem for image processing and machine learning.
Simultaneously, by simplifying creation and parsing of DICOM-compliant files, \emph{highdicom} achieves interoperability with the medical imaging systems that hold the data used to train and run ML models, and ultimately communicate and store model outputs for clinical use.
We demonstrate through experiments with slide microscopy and computed tomography imaging, that, by bridging these two ecosystems, \emph{highdicom} enables developers and researchers to train and evaluate state-of-the-art ML models in pathology and radiology while remaining compliant with the DICOM standard and interoperable with clinical systems at all stages.

To promote standardization of ML research and streamline the ML model development and deployment process, we made the library available free and open-source at
\url{https://github.com/herrmannlab/highdicom}.

%% file: sections/introduction.tex
Recent breakthroughs in machine learning (ML) and computational processing capabilities have led to the development of ML models that demonstrate unprecedented performance on a variety of highly complex computer vision tasks~\cite{pmid26017442}.
State-of-the-art convolutional neural network models now regularly achieve near-human or even superhuman performance on a variety of challenging vision tasks and across different imaging modalities, including segmentation and classification of slide microscopy images in pathology~\cite{pmid31308507,pmid33953404,pmid33990804} as well as computed tomography or magnetic resonance images in radiology~\cite{pmid31894144,pmid29944078,pmid31110349}.
Over the last couple of years, ML models have evolved technically within the research domain~\cite{pmid31753987} and the vision is that these models will soon be applied widely in clinical practice to support pathologists and radiologists in interpretation of images and ultimately improve diagnostic accuracy and efficiency~\cite{pmid31151893,pmid28447900,pmid29944078,pmid31355445}.
To realize this vision, healthcare enterprises are now tasked with evaluating model performance in clinical context and integrating the outputs of ML models into clinical workflows.
Similarly, development of models may be expedited if image annotations generated by clinical experts and stored within clinical information systems could be directly consumed by ML training and validation pipelines.
Unfortunately, this is currently impeded by the lack of standard interfaces for exchange of image annotations and ML model outputs between image analysis, image display, and image management systems.

Digital Imaging and Communication in Medicine (DICOM) is the internationally accepted standard for communication of medical images and related information across a wide range of medical imaging modalities and disciplines.
Hospitals around the world have established an extensive enterprise imaging infrastructure, workflows, and software applications based on DICOM~\cite{pmid27245774} and pathology and radiology are converging towards using DICOM for communication of digital images~\cite{pmid29619278,pmid30533276,10.4103/jpi.jpi_98_20,ihe-palm-dpia}.
However, existing pathology as well as radiology systems primarily rely on non-standard formats and interfaces for the storage and exchange of image annotations and computational image analysis results, to which we hereafter collectively refer as \emph{annotations}.
Similarly, ML models developed by researchers generally receive and return annotations in a variety of customized formats that are incompatible with clinically available image management and display systems and that lack metadata required for interpretation and use of the information in clinical context.
Instead, it would be desirable if ML models were developed according to the FAIR guiding principles~\cite{pmid26978244} using standardized metadata to allow for annotations to be findable, accessible, interoperable, and reusable.
The DICOM standard provides information object definitions (IODs), such as Segmentations and Structured Reports, for annotations~\cite{pmid27257542,pmid29092948}, and implementation of these IODs to enable interoperable storage and communication of ML model outputs has been proposed by the Integrating the Healthcare Enterprise (IHE) Radiology Technical Committee~\cite{ihe-rad-air}.

Python is the \textit{de facto} standard programming language of data science and provides a rich ecosystem for scientific computing, image processing, and machine learning~\cite{pmid32015543,pmid32939066,sklearn,skimage}.
The majority of ML models are developed and deployed in the form of Python programs.
The \emph{pydicom} library~\cite{mason2011t} provides data structures and routines for storing and accessing data of DICOM datasets (parts 5 and 6 of the DICOM standard) as well as reading and writing DICOM files (part 10 of the DICOM standard).
However, \textit{pydicom} has no concept of IODs (parts 3 and 16 of the DICOM standard) and as such leaves it to each developer to set all attributes required by an IOD manually and ensure that they follow all relevant constraints when creating new DICOM objects containing annotations.
Similarly, parsing the annotation IODs for the information relevant to a particular ML task using the \textit{pydicom} API is challenging due to their highly nested and interdependent structure.
Consequently, both tasks are slow, complex, and error-prone and require considerable knowledge of the DICOM standard.
We therefore identified a need for a higher-level abstraction layer between the ML model developer and the low-level encoding rules of the DICOM standard.
This motivated us to create the open-source \emph{highdicom} library, which provides a high-level application programming interface for creating and reading annotations in DICOM format using the Python programming language.
Our goal in releasing this library is to enable ML processes that achieve interoperability between ML models and clinical information systems throughout the entire model development and deployment lifecycle while avoiding the complexity that this currently entails.
Further, we aimed to create a library that is applicable across a range of common ML tasks and imaging domains.

In this article, we first describe the design and implementation of the \emph{highdicom} library to meet this unmet need and then assess the library's capabilities in encoding and decoding annotations (either generated by human readers or ML models) in DICOM format.
We perform experiments that demonstrate the use of the library during ML model training and inference and show how the library enables the development of ML models that are interoperable with established image management and display systems and thus can be readily integrated into an enterprise medical imaging environment.
To this end, we consider a variety of clinically-relevant computer vision problems and multiple imaging modalities across different medical disciplines, placing a focus on lung tumor detection in slide microscopy images in pathology and computed tomography images in radiology as an illustrative use case.

%% file: sections/methods.tex
\subsection*{Design Overview and Application Programming Interface (API)}

\begin{figure}[hbt]
    \includegraphics[width=\textwidth]{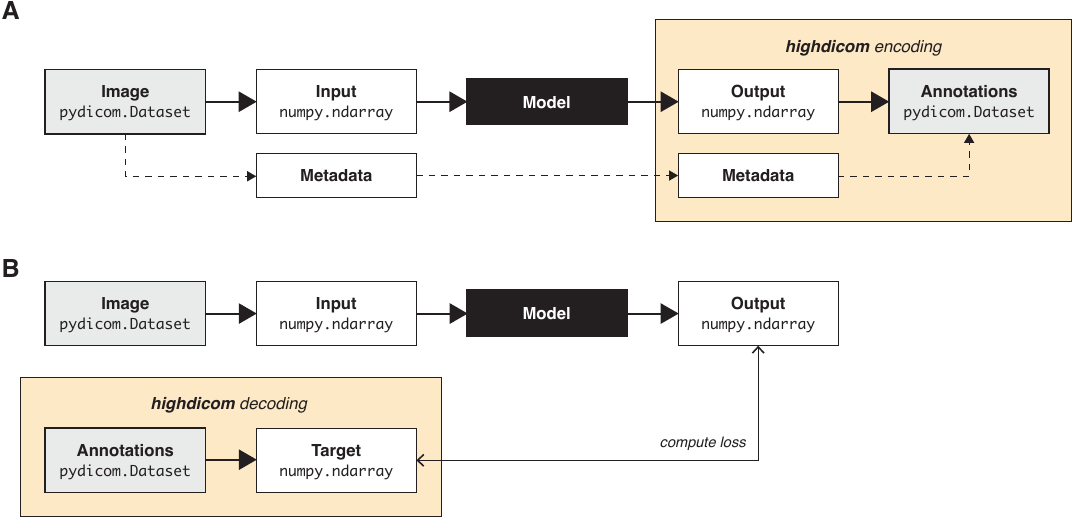}
    \caption{%
        Intended use of \emph{highdicom} in data pipelines during machine learning model training and inference workflows.
        \textbf{A}. Encoding of model outputs upon inference in the postprocessing pipeline.
        \textbf{B}. Decoding of image annotations for model training in the preprocessing pipeline.
    }\label{fig:data-pipelines}
\end{figure}

The software components responsible for transforming the data input and output from ML models, and thereby ensuring interoperability with adjacent systems, are commonly referred to as data pipelines~\cite{hapke2020,Sambasivan2021}.
During inference, pipelines are responsible for retrieving and preprocessing input images into an in-memory format that can be consumed by the model and encoding the model's in-memory outputs into a form suitable for communication and storage.
During training, they retrieve and preprocess input images and additionally, if required, decode annotations into an in-memory representation of the target for model training.
The \emph{highdicom} library is intended to operate within data pipelines that connect clinical infrastructure using the DICOM standard to popular Python ML frameworks such as \emph{PyTorch}~\cite{NIPS2019.9015} and \emph{Tensorflow}~\cite{abadi2016tensorflow}, and is focused on annotations rather than the input images themselves.
The library's core functionality is twofold:
First, encoding model outputs in form of \emph{NumPy} arrays together with relevant metadata into annotations in the form of \emph{pydicom} objects (figure~\ref{fig:data-pipelines}A).
Second, decoding annotations provided as \emph{pydicom} obtains to obtain targets in form of \emph{NumPy} arrays (figure~\ref{fig:data-pipelines}B) by reading and interpreting the included metadata.
We chose the n-dimensional \emph{NumPy} array data structure~\cite{pmid32939066} as an in-memory representation of model outputs and targets because it is interoperable with \emph{pydicom} as well as \emph{PyTorch} and \emph{Tensorflow} and many other well-established Python image processing libraries (e.g., \emph{OpenCV}~\cite{opencv} and \emph{ITK}~\cite{yoo2002}).

\paragraph*{API Overview}

\begin{figure}[hbt!]
    \includegraphics[width=\textwidth]{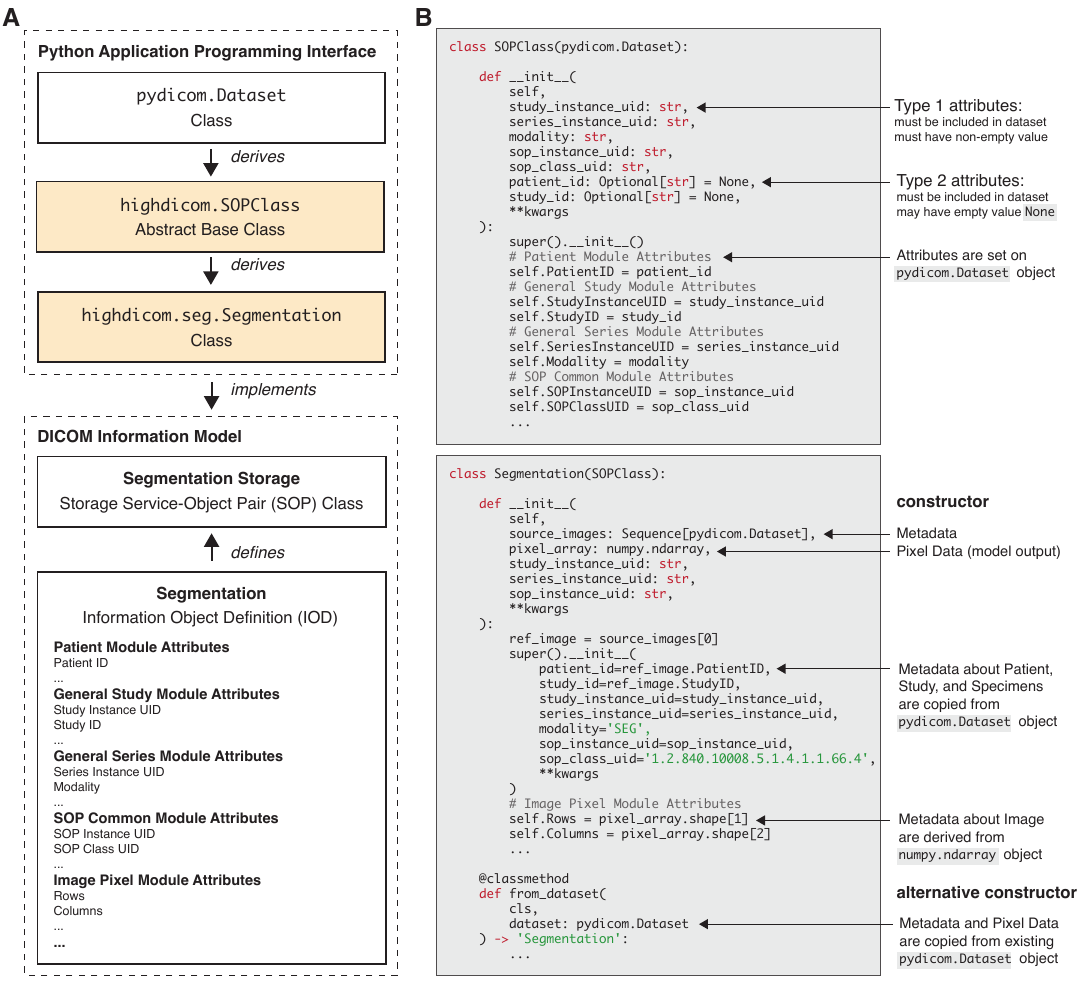}
    \caption{%
        Implementation of the DICOM Information Model in Python.
        \textbf{A}. The \emph{highdicom} Python abstract base class \texttt{highdicom.SOPClass} and its relationship to an DICOM Information Object Definition (IOD) and DICOM Storage Service-Object Pair (SOP) Class.
        \textbf{B}. A \emph{highdicom} Python class for a specific DICOM IOD and SOP Class (exemplified by \texttt{highdicom.seg.Segmentation} that implements the DICOM Segmentation Storage SOP Class defined by the DICOM Segmentation IOD).
    }\label{fig:implementation}
\end{figure}

We designed \emph{highdicom} following the object-oriented programming paradigm and modelled the API according to the DICOM Information Model, which specifies different abstract data types that are referred to as Information Object Definitions (IODs) (figure~\ref{fig:implementation}).
An IOD defines the set of required and optional DICOM attributes that may be included into DICOM objects.
We selected various IODs for storage of annotations and implemented each in \emph{highdicom} as a Python class.

Strictly speaking, each Python class implements a DICOM Storage Service-Object Pair (SOP) Class, which is the data structure within the DICOM standard that stores the attributes defined by an IOD\@.
An instance of such a Python class thus represents a DICOM SOP Instance and serves as a container for a DICOM Data Set, where each instance attribute holds the value of a DICOM Data Element.

The Python classes are ultimately derived from the \texttt{pydicom.Dataset} class from the existing \texttt{pydicom} package and therefore inherit low-level behaviors, such as accessing, setting, iterating over data elements and reading/writing to/from files that many developers are already familiar with.
It further allows developers to retain low-level control over all data elements in order to add to or alter information in objects constructed by \emph{highdicom}.
Below \texttt{pydicom.Dataset} in the class hierarchy, there is a common abstract base class called \texttt{highdicom.SOPClass} (figure~\ref{fig:implementation}A), which abstracts the attributes that are required by all SOP classes.
Specific SOP classes are then implemented by dedicated Python classes that are derived from the abstract base class (figure~\ref{fig:implementation}B).
In this way, we aim to provide an idiomatic Python interface that abstracts as much of the low-level DICOM encoding and decoding rules as possible while staying close to the standard DICOM terminology to avoid potential ambiguities.

\paragraph*{Encoding of DICOM SOP Instances}

The process of encoding information in derived objects is implemented in the constructor methods of the corresponding SOP classes (either in the \texttt{highdicom.SOPClass} abstract base class or in derived IOD-specific classes).
For construction of an SOP instance, the developer provides the image-derived information that is outputted by a model (e.g., pixel data or graphic data) together with descriptive contextual information that the standard requires for the corresponding IOD\@.
Attribute values that are static or can be derived from provided arguments are automatically set upon object construction.
For example, relevant metadata about the patient, the study, or the specimen are automatically copied from the metadata of provided source images and references to the source images are included in the derived objects (figure~\ref{fig:dicom-iods}A-B).
Furthermore, the constructor automatically validates the content of created SOP instances through runtime checks to ensure that constructed objects are fully compliant with the relevant IOD in the standard.

By design, all required information must be passed to the SOP class constructor when creating the object, and thereafter the object remains immutable through the \emph{highdicom} API (though an experienced developer may use the lower-level interface provided by the \emph{pydicom} API to modify the object if required).
This means that the constructor can validate all input parameters at once accounting for all interdependencies and conditional logic between attributes.
It also reflects the intent of the standard in that DICOM objects are immutable following creation.

\paragraph*{Decoding of DICOM SOP Instances}

The \emph{pydicom} library provides a powerful low-level Python interface to developers to access DICOM data elements of a dataset directly, with little abstraction from the details of the data format.
While this is appropriate for many image objects, the complexity of the derived objects used for annotations means that accessing the desired information using the \emph{pydicom} API requires a detailed knowledge of the underlying data structures and in our experience results in a verbose, cumbersome, and error-prone process.
Therefore, we have endowed \emph{highdicom} SOP classes with additional methods (not in the standard) that provide a means for developers to access, filter, and interpret the content of a DICOM object when preparing image annotations to be used as targets for a training algorithm.
In addition, \emph{highdicom} SOP classes implement alternative constructor methods that allow for the creation of \emph{highdicom} SOP instances from existing \texttt{pydicom.Dataset} objects, which were read from a file or retrieved over network, and thereby enhance the objects with additional, modality-specific methods and properties for data access.

\paragraph*{Data types and structures}

The majority of DICOM metadata attribute values that are passed to and returned from the \textit{highdicom} API upon encoding and decoding of SOP instances have primitive, built-in Python types such as strings (\texttt{str}), integers (\texttt{int}), and floats (\texttt{float}).
To further encapsulate closely related metadata of composite DICOM data types (DICOM Sequences or Sequence Items) and to improve code readability and reusability, the \emph{highdicom} API further provides custom Python types, which are implemented in the form of Python classes and are generally derived from either \texttt{pydicom.Dataset} or \texttt{pydicom.Sequence}.
DICOM bulkdata values such as pixel data or vector graphic data are passed to and returned from \textit{highdicom} Python classes as NumPy objects (\texttt{numpy.ndarray}).

\subsection*{Storage of annotations in DICOM format}

\begin{figure}[hbt!]
    \includegraphics[width=\textwidth]{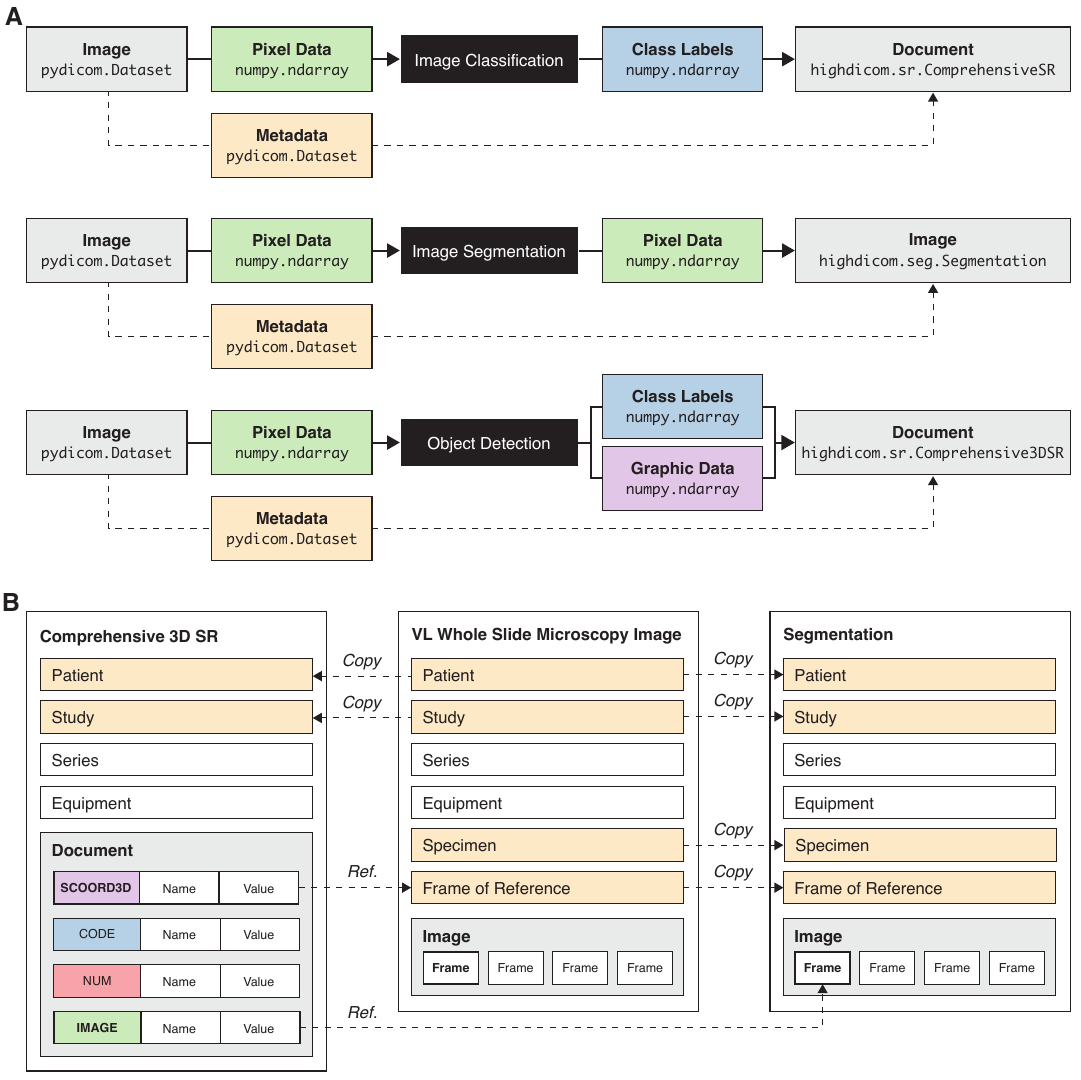}
    \caption{%
        Encoding of machine learning model outputs in DICOM\@.
        \textbf{A}. Information entities and the Python types used to represent machine learning model inputs (images) and outputs (image-derived information) for three common decision problems.
        \textbf{B}. Schematic overview of the content of source image objects (exemplified by a DICOM VL Whole Slide Microscopy Image) and derived objects (DICOM Comprehensive 3D SR and DICOM Segmentation).
        Note that descriptive metadata is copied from source to derived objects and derived objects may reference information contained in source images or other derived objects.
    }\label{fig:dicom-iods}
\end{figure}

Having described the general approach taken by our library, we now begin to discuss the individual IODs that we selected for implementation.
The DICOM standard specifies a wide range of IODs for different types of DICOM objects, including images acquired by various modalities (e.g., computed tomography or whole slide microscopy) as well as image-derived information generated by image display, processing, or analysis systems~\cite{pianykh2008}.
For implementation in the \emph{highdicom} library, we considered standard IODs that provide mechanisms to store image annotations for common ML tasks across pathology and radiology use cases.
We thereby focused on the following decision problems and their corresponding annotations (figure~\ref{fig:dicom-iods}A)~\cite{pmid26017442}:
\begin{enumerate}
    \item \emph{Image classification} --- class labels in the form of discrete binary or categorical values and optionally class scores in the form of continuous probabilistic values (figure~\ref{fig:dicom-iods}A upper panel)
    \item \textit{Image segmentation} --- class labels at pixel resolution that identify semantically distinct regions of interest (ROIs) within an image in the form of raster graphics (figure~\ref{fig:dicom-iods}A middle panel)
    \item \emph{Object detection} --- spatial coordinates for individual ROIs in the form of vector graphics (commonly bounding boxes), combined with class labels and detection scores (figure~\ref{fig:dicom-iods}A lower panel)
\end{enumerate}

We identified three IODs that together allow for the encoding of annotations for these common use cases: the Segmentation IOD and two Structured Report (SR) IODs.
The Segmentation IOD was selected to encode ROIs returned by image segmentation models as raster graphics.
The Comprehensive SR and Comprehensive 3D SR IODs were chosen to encode vector graphic ROIs returned by object detection models as well as class labels, scores, and measurements returned by image classification and regression models (figure~\ref{fig:dicom-iods}A).
All three IODs are designed to be agnostic of the imaging modality and able to support use cases across medical disciplines including pathology and radiology.

\paragraph*{DICOM Segmentation images}

The Segmentation IOD is implemented in \texttt{highdicom} as the \texttt{highdicom.seg.Segmentation} Python class and allows for the encoding of one or more components, which in DICOM are referred to as \emph{segments}.
Each segment may represent a pixel class (category) or an individual instance of a given class as generated by semantic segmentation or instance segmentation models~\cite{hafiz2020}, respectively.
Segments may further have binary or fractional type, either representing a mask of Boolean values where non-zero pixels encode class membership or a mask of decimal numbers where pixels encode class probability.

In order to encode a DICOM Segmentation image, the developer passes to the constructor a mask as a \texttt{numpy.ndarray} (of either Boolean, integer, or floating point data type) along with additional metadata that describe the meaning of each segment within the segmentation (\texttt{highdicom.seg.SegmentDescription}) and the algorithm responsible for producing the segmentation (\texttt{highdicom.AlgorithmIdentificationSequence}).

To facilitate decoding of DICOM Segmentation images, the \texttt{highdicom.seg.Segmentation} class provides methods that allow developers to filter segments by their label, segmented property category or type, or tracking identifiers.
It further provides methods to obtain a segmentation mask as a \texttt{numpy.ndarray} for a given set of segments and source image frames.
While conceptually straightforward, in practice several steps are necessary to achieve this correctly:
\begin{enumerate*}[label=(\roman*)]
    \item{determining which frames stored in the Segmentation image are relevant to a given set of segments and source image frames based on the multi-frame dimension indexing information},
    \item{sorting the Segmentation image frames according the query},
    \item{adding in missing pixel values in case of sparse Segmentation images where background image frames were omitted during encoding to save storage space}
	\item{(optionally) combining multi binary segments into a multi-class label map}.
\end{enumerate*}

\paragraph*{DICOM Structured Report documents}

\begin{figure}[hbt!]
    \includegraphics[width=\textwidth]{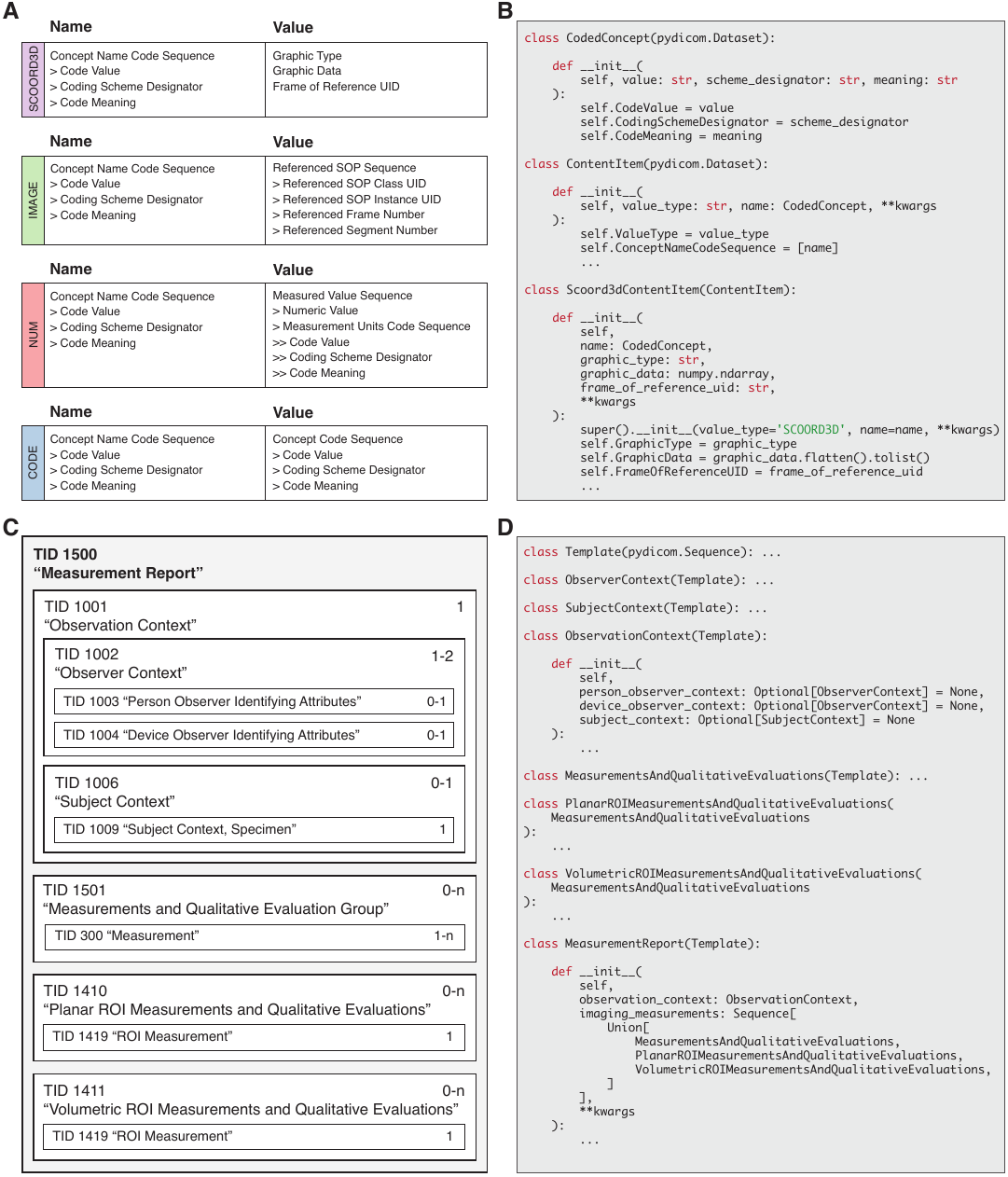}
    \caption{%
        Encoding of annotations as DICOM Structured Reporting (SR) content items and templates for inclusion into an SR document.
        \textbf{A}. SR content items of different values types.
        \textbf{B}. Implementation of SR content items in \emph{highdicom} by classes that inherit from \texttt{pydicom.Dataset}.
        \textbf{C}. SR template TID 1500 ``Measurement Report'' and included sub-templates.
        \textbf{D}. Implementation of SR templates in \emph{highdicom} by classes that inherit from \texttt{pydicom.Sequence}.
    }\label{fig:dicom-sr}
\end{figure}

There are various IODs defined by the standard that utilize structured reporting, but we selected the Comprehensive SR (\texttt{highdicom.sr.ComprehensiveSR}) and Comprehensive 3D SR (\texttt{highdicom.sr.Comprehensive3DSR}) IODs for implementation in \emph{highdicom} because they provide the most flexible mechanisms for storing annotations.
In addition to the IOD definitions, the standard provides SR templates, which serve as schemas that define how the content of an SR document shall be structured and how the information shall be encoded.
A template consists of a sequence of \emph{content items}, each defining a name-value pair (or question-answer pair) that encodes a domain-specific property or concept (figure~\ref{fig:dicom-sr}A).
Notably, both concept names and values have a composite data type and are each encoded by one or more DICOM attributes.
Concept names are coded using standard medical terminologies and ontologies such as the DICOM Controlled Terminology or the Systematized Nomenclature of Medicine Clinical Terms (SNOMED CT) and thereby get endowed with an explicit, domain-specific meaning~\cite{pmid9865038}.
The structure of the corresponding value depends on the \emph{value type}, which defines a set of DICOM attributes that are included in the SR document to represent the assigned value.

Within \emph{highdicom}, the \texttt{highdicom.sr.CodedConcept} class is an important data type that encapsulates the DICOM attributes required to code a concept using a standard coding scheme within a single Python object.
We further contributed lower-level data types to the underlying \emph{pydicom} library that provide programmatic access to codes included in the DICOM standard, specifically the DICOM Controlled Terminology (DCM), SNOMED-CT (SCT), and Unified Code for Units of Measure (UCUM) coding schemes.
These codes that are included in the \emph{pydicom} library are fully compatible with the coded concepts of the \emph{highdicom} library and can generally be used interchangeably throughout the API\@.
Furthermore, for each of the different DICOM content item value types we have implemented a separate Python class that is derived from \texttt{pydicom.Dataset} and encapsulates both the coded concept name and the corresponding value of the given type (figure~\ref{fig:dicom-sr}B).

Notable content item classes include \texttt{highdicom.sr.CodeContentItem}, which may be used to store class labels as coded values, and \texttt{highdicom.sr.NumContentItem}, which may be used to store a measurement along with its unit.
ROIs may be either encoded by value or by reference and stored within or outside of the SR document content, respectively.
In the case of vector graphics (including but not limited to bounding boxes), the graphic data may be stored within the SR document and encoded via DICOM content items of value type \texttt{SCOORD3D}, which encodes 3D spatial coordinates of geometric objects in the frame of reference (patient or slide coordinate system).
This value type is implemented in \emph{highdicom} by the \texttt{highdicom.sr.Scoord3DContentItem} Python class (figure~\ref{fig:dicom-sr}B).
In the case of raster graphics, the pixel data of Segmentation images are stored outside of the SR document, but specific segments can be referenced from within the SR document via content items of value type \texttt{IMAGE}. (implemented by the \texttt{highdicom.sr.ImageContentItem} Python class), which includes DICOM identifiers for the referenced image object and segments contained therein.

The standard provides different SR templates for a variety of common clinical use cases and diagnostics tasks, such as recording X-Ray dose exposure or reporting echocardiography findings.
We chose to implement the more generic template TID 1500 ``Measurement Report'' in \emph{highdicom} for encoding annotations, because the template provides standard content items to describe measurements and qualitative evaluations of images as well as individual image ROIs (figure~\ref{fig:dicom-sr}C) and because it has already been successfully used for standardized communication of quantitative image analysis results~\cite{pmid27257542,pmid29092948}.
Importantly, sub-templates that can be included in TID 1500 to allow for encoding of annotations (figure~\ref{fig:dicom-sr}C).
Within the library's API, these selected templates are implemented by Python classes, which are derived from an abstract base class \texttt{highdicom.sr.Template}, which is in turn derived from \texttt{pydicom.Sequence} (figure~\ref{fig:dicom-sr}D).
The constructors of these Python classes require the developer to pass the relevant data via named parameters but then handle its inclusion in the template with the correct concept names and as well as ensuring all constraints are satisfied.

When decoding SR documents, the high degree of nesting in the document tree and the variable order of content items at each level, means that finding a particular content item of interest in the tree potentially requires multiple nested loops.
Further, as described above, each content item is a collection of data elements that must first be parsed and interpreted as a unit.
The Python classes that implement SR templates and individual SR content items provide methods and properties to facilitate data access.
Using the provided methods, measurement groups within a \texttt{highdicom.sr.ComprehensiveSR} or \texttt{highdicom.sr.Comprehensive3DSR} object can be filtered by their finding type, finding site, or tracking identifiers.
Individual measurements and qualitative evaluations contained within these groups can similarly be filtered by their concept name.
Furthermore, \emph{highdicom} classes representing SR templates and content items provide access to their content items or values, respectively, through Python properties that return the data either as a built-in Python type or a custom \textit{highdicom} type (which will typically match the type of the argument passed to the constructor).

%% file: sections/results.tex
Having laid the foundation through the description of the library's design and implementation, we now proceed to demonstrating the capabilities of the library.
We consider a concrete use case of developing machine learning models for lung tumor detection in both pathology and radiology and deploying the models clinically using a common platform and framework that is applicable independent of the medical discipline or imaging modality.
In this section, we first describe the steps necessary to encode the annotations in DICOM using \emph{highdicom}, including the description of the detected region of interest, the identified finding, and related measurements and qualitative evaluations.
We then show through a series of experiments how \emph{highdicom} can streamline ML model training and inference for this use case.

\subsection*{Highdicom facilitates encoding of image annotations in DICOM format}

\paragraph*{Structured reporting using standard medical terminologies}

While the approach of using standardized vocabularies is powerful and important for interoperability, it complicates working with the data.
For example, comparing two concepts for equality requires comparison of their code values, coding scheme designators, and coding scheme versions.
The \texttt{highdicom.sr.CodedConcept} and the lower-level \emph{pydicom} types facilitate the use of coded concepts for structured reporting of annotations in Python at a high level of abstraction (code snippet~\ref{code:concepts}).

\begin{lstlisting}[label={code:concepts},caption={Coding of concepts using domain-specific terminologies. ``Neoplasm'' and ``Tumor'' are synonyms that map to the same code within the SNOMED-CT coding scheme.}]
import highdicom as hd
from pydicom.sr.codedict import codes

tumor = hd.sr.CodedConcept("108369006", "SCT", "Tumor")
neoplasm = codes.SCT.Neoplasm

assert(tumor == neoplasm)  # True
assert(tumor.value == neoplasm.value)  # True
assert(tumor.scheme_designator == neoplasm.scheme_designator)  # True
\end{lstlisting}

\paragraph*{Describing ROI evaluations and measurements}

The Coded Concept type forms the basis for additional higher-level composite data types for DICOM structured reporting such as SR content items.
Code snippet~\ref{code:content-items} demonstrates example content items for the encoding of a tumor image region of interest, the tumor finding, and an associated tumor measurement.

\begin{lstlisting}[label={code:content-items},caption={Encoding of regions of interest, qualitative evaluations and measurements.}]
import numpy as np
import highdicom as hd
from pydicom.sr.codedict import codes

# Encode bounding box enclosing the tumor region using 3D spatial coordinates
region_item = hd.sr.ImageRegion3D(
    graphic_type=hd.sr.GraphicTypeValues3D.POLYGON,
    graphic_data=np.array([
        [45., 34., 0.],
        [49., 34., 0.],
        [49., 42., 0.],
        [45., 42., 0.],
        [45., 34., 0.]
    ]),
    frame_of_reference_uid="1.2.3.4"
)

# Encode a qualitative evaluation describing the "morphology" as "adenocarcinoma"
morphology_item = hd.sr.QualitativeEvaluation(
    name=codes.SCT.AssociatedMorphology,
    value=hd.sr.CodedConcept(
        value="8551/3",
        scheme_designator="ICDO3",
        meaning="Acinar adenocarcinoma"
    )
)

# Encode a measurement with name "area" and of unit "square millimeter"
area_item = hd.sr.Measurement(
    name=codes.SCT.Area,
    value=12.07,
    unit=codes.UCUM.SquareMillimeter
)
\end{lstlisting}

This demonstrates using the SCT vocabulary built in to \emph{pydicom} to encode a concept name as `Morphology', and a domain-specific coding scheme, the International Classification of Diseases for Oncology (ICD-O), to specify the exact type of tumor as the concept value.
Of note, the area measurement in our example is encoded in a well-defined physical unit, as would be expected for clinical decision making.
The corresponding image region is defined in the same physical space.
In DICOM, image regions may be defined by spatial coordinates within either the pixel matrix of an individual image, or, as in this example, the frame of reference (the 3D patient- or slide-based physical coordinate system).
While the former appears more straightforward, the latter is more general and allows for annotations derived from transformed versions of the original images with arbitrary affine transformations (rotations, scaling, etc.) as well as crops.

\paragraph*{Creation of DICOM annotation objects}

The computer vision problem of tumor detection could be solved using either an object detection or image segmentation model.
Accordingly, the output of these models and the annotations used to train them can be encoded using the \texttt{highdicom.sr.Comprehensive3DSR} (code snippet~\ref{code:object-detection}) and \texttt{highdicom.seg.Segmentation} (code snippet~\ref{code:segmentation}) classes respectively.
In either case, this involves describing the finding and the anatomical site of the finding as well as supplying relevant contextual metadata such as the device or person reporting the observation.
However, note that it is not necessary to specify patient, study, or specimen information since \emph{highdicom} copies this metadata directly from the source images provided as evidence to the constructor.

\begin{lstlisting}[label={code:object-detection},caption={Creation of a DICOM Comprehensive 3D SR document instance to encode predictions of an object detection model using content items constructed in snippet~\ref{code:content-items}.}]
import highdicom as hd
from pydicom.dataset import Dataset
from pydicom.sr.codedict import codes

def load_image() -> Dataset: ...

source_image = load_image()

# Describe the device that is reporting the observation
observation_context = hd.sr.ObservationContext(
    observer_device_context=hd.sr.ObserverContext(
        observer_type=codes.DCM.Device,
        observer_identifying_attributes=hd.sr.DeviceObserverIdentifyingAttributes(
            uid=hd.UID(),
            name="GPU01"
        )
    )
)
# Describe the region of interest finding and associated measurements and evaluations
roi_description = hd.sr.PlanarROIMeasurementsAndQualitativeEvaluations(
    finding_type=codes.SCT.Neoplasm,
    finding_sites=[hd.sr.FindingSite(anatomic_location=codes.SCT.Lung)],
    referenced_region=region_item,
    measurements=[area_item],
    qualitative_evaluations=[morphology_item]
)
# Construct a measurement report from the ROI(s)
measurement_report = hd.sr.MeasurementReport(
    observation_context=observation_context,
    procedure_reported=codes.SCT.ImagingProcedure,
    imaging_measurements=[roi_description]
)
# Construct an SR document containing the measurement report
report_document = hd.sr.Comprehensive3DSR(
    evidence=[source_image],
    content=measurement_report[0],
    series_number=1,
    series_instance_uid=hd.UID(),
    sop_instance_uid=hd.UID(),
    instance_number=1,
    manufacturer="MGH Computational Pathology",
    manufacturer_model_name="Object Detection Model",
    software_version="v1",
    device_serial_number="XYZ"
)
\end{lstlisting}

\begin{lstlisting}[label={code:segmentation},caption={Creation of a DICOM Segmentation image instance to encode predictions of a semantic image segmentation model.}]
from typing import Tuple

import highdicom as hd
import numpy as np
from pydicom.sr.codedict import codes
from pydicom.dataset import Dataset

def load_image() -> Tuple[Dataset, np.ndarray]: ...

def segment_image(image: np.ndarray) -> np.ndarray: ...

metadata, frames = load_image()
probabilities = segment_image(frames)

segment_description = hd.seg.SegmentDescription(
    segment_number=1,
    segment_label="Tumor",
    segmented_property_category=codes.cid7150.MorphologicallyAbnormalStructure,
    segmented_property_type=codes.cid7159.Neoplasm,
    algorithm_type=hd.seg.SegmentAlgorithmTypeValues.AUTOMATIC,
    algorithm_identification=hd.AlgorithmIdentificationSequence(
        name="highdicom-experiments",
        version="v1.0",
        family=codes.cid7162.ArtificialIntelligence
    )
)
segmentation_image = hd.seg.Segmentation(
    source_images=[metadata],
    pixel_array=probabilities,
    segmentation_type=hd.seg.SegmentationTypeValues.FRACTIONAL,
    segment_descriptions=[segment_description],
    series_instance_uid=hd.UID(),
    series_number=1,
    sop_instance_uid=hd.UID(),
    instance_number=1,
    manufacturer="MGH Computational Pathology",
    manufacturer_model_name="Segmentation Model",
    software_versions="v1",
    device_serial_number="Server XYZ"
)
\end{lstlisting}

\subsection*{Highdicom facilitates efficient loading and decoding of images and corresponding annotations}

When it comes to training a model for tumor detection, annotations may be provided in the form of either raster graphics within a Segmentation image, or vector graphics within an SR document.
In both cases, \emph{highdicom} provides methods that simplify access to, and interpretation of, the relevant content in the annotation SOP instances.
If annotations are provided as raster graphics within a Segmentation image, model training may require combining binary bit planes from multiple segments in the Segmentation image to create a single label map, represented as a NumPy array, in which pixels encode tumor identities.
If instead annotations are provided as vector graphics within an SR document, the spatial coordinates of image regions will need to be collected from within the document content tree and passed as NumPy arrays to training processes.
Snippets~\ref{code:parse-sr} and~\ref{code:parse-seg} show example usage of the methods that \emph{highdicom} provides for these purposes.

\begin{lstlisting}[label={code:parse-sr},caption={Example of parsing a DICOM SR document to find lung tumor regions.}]
from typing import List
import highdicom as hd
import numpy as np
from pydicom import dcmread
from pydicom.sr.codedict import codes

# Create highdicom SR document object from a dataset read from file
dataset = dcmread('annotation.dcm')
sr_document = hd.sr.Comprehensive3DSR.from_dataset(dataset)

# Ensure that the document content is a TID 1500 "Measurement Report"
assert isinstance(sr_document.content, hd.sr.MeasurementReport)

# Find bounding polygons for lung tumor regions of interest
tumor_regions: List[np.ndarray] = [
    group.roi.value
    for group in sr_document.content.get_planar_roi_measurement_groups(
        finding_type=codes.SCT.Neoplasm,
        finding_site=codes.SCT.Lung,
        graphic_type=hd.sr.GraphicTypeValues3D.POLYGON
    )
]
\end{lstlisting}

\begin{lstlisting}[label={code:parse-seg},caption={Example of parsing a DICOM Segmentation image to create a label map of lung tumor regions for a given source frame.}]
from typing import List
import highdicom as hd
import numpy as np
from pydicom import dcmread
from pydicom.sr.codedict import codes

# Create highdicom segmentation object from a dataset read from file
dataset = dcmread('annotation.dcm')
seg_image = hd.seg.Segmentation.from_dataset(dataset)

# Find numbers of segments for which the segmented property is "tumor"
tumor_segment_numbers: List[int] = seg_image.get_segment_numbers(
    segmented_property_type=codes.SCT.Neoplasm
)

# Get label map with these segments for a given SOP instance
label_map: np.ndarray = seg_image.get_pixels_by_source_instance(
    source_sop_instance_uids=['1.2.3.4'],
    segment_numbers=tumor_segment_numbers,
    combine_segments=True,
    relabel=True
)
\end{lstlisting}

\subsection*{Highdicom facilitates decoding and encoding of annotations during model training and inference, respectively}

To establish a proof-of-concept standard-based ML workflow and to demonstrate the utility of the \emph{highdicom} library for ML, we performed a set of experiments on the training and evaluation of deep convolutional neural network (CNN) models using publicly-available slide microscopy (SM) and computed tomography (CT) image data sets.
We emphasize that our intent is to demonstrate a complete ML workflow for pathology and radiology fully based on DICOM, rather than create models with optimal performance or reach state-of-the-art for a particular task.

For pathology, we trained and evaluated models using lung cancer collections of slide microscopy (SM) images from The Cancer Imaging Archive (TCIA)~\cite{pmid23884657} that were acquired as part of The Cancer Genome Atlas (TCGA) Lung Adenocarcima (LUAD) or Lung Squamous Cell Carinoma (LUSC) projects and which we converted into DICOM format as previously described~\cite{pmid30533276,pmid31057981}.
For radiology, we used the collection of CT images of the Lung Image Database Consortium (LIDC) and Image Database Resource Initiative (IDRI) (LIDC-IDRI)~\cite{Armato2011Lidc,Armato2015Lidc,pmid23884657}, which were already available in DICOM format.
We used available measurements and qualitative evaluations for these SM and CT files provided by TCIA as image annotations, which we encoded in DICOM SR documents or DICOM Segmentation images using \emph{highdicom} (see supplementary methods), resulting in training sets for CT lung nodule detection and SM image classification encoded entirely within DICOM format.

\begin{figure}[h]
    \includegraphics[width=\textwidth]{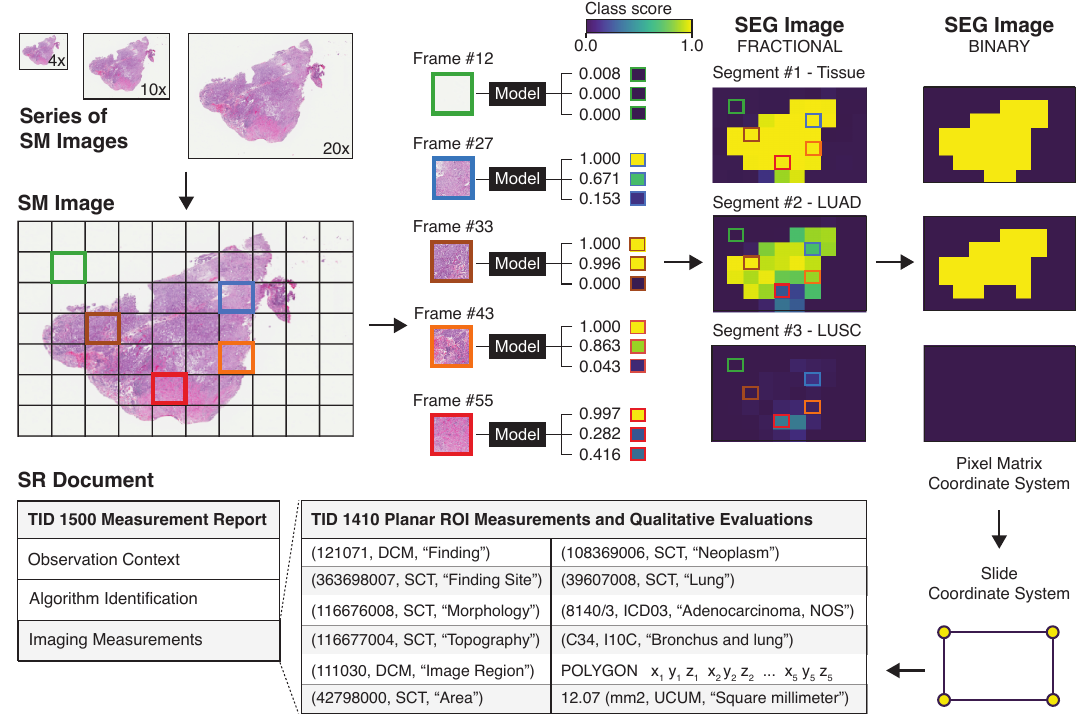}
    \caption{%
        Schematic overview of output post-processing pipelines of the pathology model, which classifies individual image frames of a multi-frame SM image of a lung tissue section specimen.
        Outputted scores get transformed into a segmentation mask from which bounding boxes of the tumor regions are derived.
        The coordinates of the bounding box vertices are stored as 3D spatial coordinates in the reference slide coordinate system.
    }\label{fig:sm-model}
\end{figure}

We developed proof-of-concept ML models based on published algorithms and implemeted data pre- and postprocessing pipelines for each model to load model inputs from DICOM SM or CT image instances, annotations from DICOM SR documents or DICOM Segmentation images respectively, and store outputs to DICOM SR instances.
For pathology, we implemented a weakly-supervised image classification model using multiple instance learning with the objective to classify individual SM image frames of lung tissue sections into slide background, normal lung tissue, lung adenocarcinoma, or lung squamous cell carcinoma similar to prior work described by Coudray et al~\cite{pmid30224757}.
To this end, we used a modified version of a ResNet-101 model~\cite{resnet}, which we initialized with parameters from pre-training on ImageNet~\cite{imagenet} and further optimized using SM image frames and image annotations from the TCGA collections similar to the algorithms described by Lerousseau et al~\cite{10.1007/978-3-030-59722-1.45} and Lu et al~\cite{pmid33649564}.
During training, each training sample was created by selecting one or more frames of an SM image from a given series (i.e., digital slide) together with the corresponding image-level annotations obtained from the SR document using \emph{highdicom}.
During inference, the data postprocessing pipeline collects predicted class probabilities for each image frame, constructs low-resolution probabilistic segmentation mask for each class (with pixels representing class probabilitisties for individual frames), and finally encodes the constructed masks in a DICOM Segmentation image with \texttt{FRACTIONAL} Segmentation Type and \texttt{PROBABILITY} Fractional Segmentation Type (figure~\ref{fig:sm-model} upper panel).
The postprocessing pipeline further thresholds the individual class probability predictions to generate a binary segmentation mask for each class (normal lung tissue, lung adenocarcinoma, or lung squamous carcinoma), performs a connected component analysis and border following to find the contours of ROIs representing class instances, and encodes each detected ROI together with additional measurements and qualitative evaluations in a DICOM Comprehensive 3D SR document (figure~\ref{fig:sm-model} lower panel).

\begin{figure}[h]
    \includegraphics[width=\textwidth]{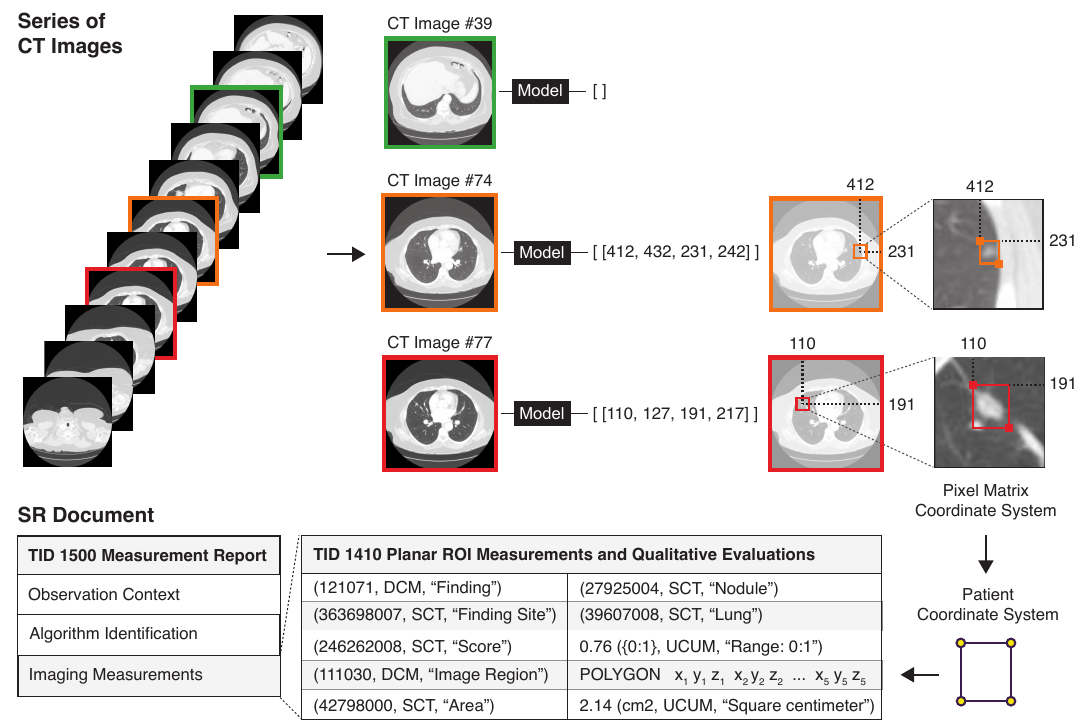}
    \caption{%
        Schematic overview of output post-processing pipeline of the radiology model, which detects lung nodules in image frames of single-frame CT images of the thorax and outputs bounding boxes of lung nodule regions.
        The coordinates of the bounding box vertices are stored as 3D spatial coordinates in the reference patient coordinate system.
    }\label{fig:ct-model}
\end{figure}

For radiology, we implemented an object detection model to detect lung nodules in individual CT slices of the chest.
We used an off-the-shelf implementation of the widely-used RetinaNet convolutional neural network~\cite{Lin2017focal} available with the \emph{torchvision} package\footnote{\url{https://pytorch.org/vision/0.8/index.html}}.
Specifically, we used a RetinaNet model with the ResNet-50 backbone~\cite{resnet} and initialized the model with weights from pre-training on the ImageNet dataset~\cite{imagenet}.
During training, each training sample was created by selecting a random CT image frame (2D axial slice) from a given series.
The annotations encoded in DICOM Segmentation images were read using \emph{highdicom} and the bounding box containing each nodule in the slice was calculated on-the-fly from the contained segments and used as a ground truth label for supervised training of the RetinaNet model.
The post-processing pipeline for the chest CT model collected predicted bounding boxes and their detection scores outputted by the RetinaNet model for every frame in the CT series and encoded them in a DICOM Comprehensive 3D SR document, with vector graphics used to represent bounding box coordinates and detection scores encoded as a measurement of the region represented by the bounding box (figure~\ref{fig:ct-model}).

\paragraph*{Annotations generated by highdicom can be stored in image management systems using DICOMweb services and visualized using DICOM-compliant display systems}

After model training, we selected one pathology and radiology model for further clinical evaluation and deployed it into a production-like environment, consisting of an image management system (IMS) with a DICOMweb interface~\cite{dicomweb} and DICOM-compliant image display systems.
Specifically, a dcm4chee-arc-light archive~\footnote{\url{https://github.com/dcm4che/dcm4chee-arc-light}} served as the IMS and we stored SM and CT images in the IMS via DICOMweb RESTful services using the \emph{dicomweb-client} Python library~\cite{pmid30533276}~\footnote{\url{https://github.com/mghcomputationalpathology/dicomweb-client}}.
Upon inference, the data preprocessing pipelines retrieved DICOM SM or CT images from the IMS over network using the \emph{dicomweb-client}, read and interpreted the image metadata and pixel data using \emph{pydicom}, and passed the pixel data as inputs to the model as NumPy arrays.
Model outputs received as NumPy arrays were encoded as DICOM SR documents in the data postprocessing pipeline using \emph{highdicom} and stored back in the IMS over network using the \emph{dicomweb-client}.

For radiology, we visualized the ground truth lung nodules using the OHIF~\footnote{\url{https://github.com/ohif/viewers}} viewer, which retrieved the DICOM Segmentation images over network using the \emph{dicomweb-client} library and displayed each segment as a raster graphic on top of the corresponding CT images (supplementary figure S3B).
We additionally visualized detected ROIs using the open-source 3D Slicer~\footnote{\url{https://slicer.org}} software (supplementary figure S3B).

For pathology, we visualized detected lung tumor regions using the Slim~\footnote{\url{https://github.com/herrmannlab/slim}} viewer, which retrieved the DICOM SR documents over network using the \emph{dicomweb-client} JavaScript library~\footnote{\url{https://github.com/dcmjs-org/dicomweb-client}} and displayed the spatial coordinates of each ROI contained in the SR documents as a vector graphic on top of the corresponding SM images (supplementary figure S4).

%% file: sections/discussion.tex
The main contributions of this paper are:
\begin{enumerate*}[label= (\roman*)]
    \item The demonstration that image annotations can be encoded and exchanged in DICOM format using existing DICOM IODs and services, respectively.
    \item The development of a software library that provides a high-level application programming interface (API) for the Python programming language to facilitate creation of DICOM objects for storage of image-derived information, including image annotations, as well as accessing and interpreting information stored in DICOM objects.
    \item The establishment of a standard-based workflow for ML model training and inference that is generally applicable across different imaging modalities, computer vision problems, and medical disciplines.
\end{enumerate*}

In developing the \emph{highdicom} library and establishing an ML workflow based on DICOM, we made several observations that merit further discussion.

\subsection*{Clinical use of machine learning model outputs in pathology and radiology requires domain-specific metadata}

Medical images and image annotations must not only contain the actual data, such as the pixel data in case of an image, but require additional metadata that enable interpretation and use of the data.
Such metadata can be grouped into information related to data representation, information about the data acquisition process and equipment, and information related to the clinical context in which the data was acquired, including identifying and descriptive information about the patient, study, and specimens.
This contextual information that describes how the data relates to the real world is crucial for unambiguous interpretation of medical images as well as any regions of interest, measurements, or qualitative evaluations derived from them.

To ensure that clinical decisions based on this information are made for the right patient and specimen and in the correct clinical setting, real-world entities need to be uniquely identifiable throughout the digital workflow.
As such it is desirable to establish an unambiguous association between the digital information (images and image annotations) on the one hand and clinically relevant real-world entities (patients, specimens, etc.) on the other hand by including clinical identifiers into digital objects.
This furthermore facilitates exchange of information between departments and institutions upon transfer and referral of patients.
DICOM specifies standard information object definitions and attributes to store and exchange digital images and image-derived information together with the relevant clinical identifiers as composite objects.
The \emph{highdicom} API facilitates access to and creation of such standard DICOM objects using the Python programming language and thereby enables data engineers and scientists to develop ML models and systems that can receive inputs and return outputs that include relevant identifiers for clinical application.
Additionally, in many cases including patient information and references to the source images, \emph{highdicom} will find and copy the relevant metadata from the dataset of the source image to reduce the room for human error as far as possible.

In addition to identifiers, DICOM objects contain descriptive metadata about the imaging target (patient or specimen), the imaging modality and procedure, the anatomical location of the imaging or surgical procedure, and in case of pathology the preparation of the specimen.
This information can be critical for the interpretation of images or image annotations by ML systems during model training or inference as well as by other systems that use or interpret model outputs.
Most importantly this descriptive metadata allows automated systems to decide whether or not a given information object may be appropriate to use in the context of the intended use or select one of several available objects for analysis or display~\cite{pmid31950302}.
Descriptive metadata is also useful for performing model validation and error analysis to determine groups of inputs, according to patient demographic information, pre-analytic specimen preparation variables, or image acquisition parameters, upon which models are under-performing.
Furthermore, the DICOM standard provides mechanisms for describing the image analysis algorithm (name, version, etc.) as well as the completeness or validity of analysis results at various stages of the clinical decision making process.
For example, the DICOM SR IODs include attributes that allow clinical users to verify or, if necessary, complete or correct ML model outputs, to record the verification or modification activity, and to create an audit trail that establishes the relationship between the document containing the verified or modified content and the predecessor document containing the unverified model outputs.
These mechanisms are critical for safe clinical application of ML models, since their outputs are generally intended for clinical decision support rather than independent decision making~\cite{Ammenwerth2019} and thus require review by a clinical expert before inclusion into the medical record.

The \emph{highdicom} library enables developers to access relevant descriptive information in received DICOM objects upon preprocessing or include such information into generated DICOM object upon post-processing and thereby make it available to downstream clinical systems.
The high-level and well-tested abstractions provided by \emph{highdicom} allow developers to achieve this goal with only a few lines of Python code.

\subsection*{Standard coding schemes enable unambiguous interpretation of image annotations}

Subtle differences in the description of imaging findings can lead to drastically different treatment decisions.
To ensure that image annotations can be interpreted unambiguously by both clinicians and devices or automated systems that may act upon the information, the terms used to describe and report annotations need to be well-defined.
DICOM structured reporting uses codes of established clinical terminologies and ontologies to describe image-derived information rather than using free text.
For example, while many words in English and other languages may be used to refer to a ``tumor'' as the finding type of the ROI, the concept can be unambiguously represented across languages and domains by the SNOMED-CT code ``108369006''.
The use of structured reports and standardized codes facilitates interpretation of image annotations by both humans and machines and is therefore critical for enabling structural and semantic interoperability between ML models and clinical systems.
The standard-based approach further facilitates the re-use of data beyond the scope of the project or use case for which they were initially created.
While there are several advantages to using codes, they are cumbersome to work with and increase the complexity of ML programs and are thus in our experience often frowned on by developers.
The \emph{highdicom} library provides data structures and methods that abstract the codes and significantly simplify using and operating on coded concepts.

While codes chosen from well-established coding schemes can significantly improve interoperability, the choice of the appropriate code can still pose a significant challenge to both developers and clinical experts.
The \emph{highdicom} library does not (and cannot) fully solve this problem. Indeed, in practice it may be the case that no standard coded concept accurately describes the annotation and a custom coding scheme is required. DICOM allows, and \emph{pydicom} and \emph{highdicom} support, the definitition of such custom coding schemes with the convention of a prefix of ``99`` followed by an identifying text string. Consumers of custom coded concepts should detect this condition and seek out-of-band information for correct interpretation of the annotations.
However, for a large range of common clinical use cases the library (together with the underlying \emph{pydicom} library) exposes value sets defined in the DICOM standard via abstractions, and by depending on these abstractions throughout its API, encourages developers to choose codes from these predefined sets.

\subsection*{Encoding image regions in a well-defined coordinate system in three-dimensional physical space allows for clinically-actionable measurements}

Establishing an unambiguous spatial relationship between ROIs and their corresponding source images for display or computational analysis requires a common frame of reference, which defines the coordinate system to uniquely localize both images or image regions with respect to the imaging target (the specimen in pathology or the patient in radiology) with both position and orientation.
Many applications simply specify ROIs relative to the pixel matrix of an image in pixel units.
However, this simple approach is problematic for interoperability, because the image pixel grid forms an ill-defined coordinate system and the location (offset, rotation, and scale) of an image with respect to the imaging target changes upon spatial transformation of the image.
DICOM specifies a frame of reference for both slide-based and patient-based coordinate systems, which enables accurate and precise localization of a ROI with respect to the patient or the specimen on the slide independent of whether affine transformations have been applied to images.
Defining ROIs in physical space in millimeter units further has the advantage that spatial ROI measurements such as diameter or area can be readily taken in this frame of reference without the need to transform coordinates, a process that can be error prone and result in incorrect measurements with potentially serious clinical implications.
The \emph{highdicom} library enables developers to work with both 2D pixel matrix and 3D frame of reference coordinates and provides developers methods to readily convert coordinates between the different coordinate systems.

\subsection*{Scaling to large numbers of image annotations in the context of slide microscopy imaging in pathology}

As demonstrated in this paper, encoding of ROIs in SR documents works for both pathology and radiology.
However, the deeply nested structure of SR documents does not scale well to object detection problems in pathology, where millions of cells or nuclei may be detected per whole slide image.
To address this challenge, DICOM Working Group 26 Pathology (WG-26) has developed a supplement for the DICOM standard that proposes the introduction of a \emph{Microscopy Bulk Simple Annotations} IOD and Annotation (ANN) modality specifically designed for the storage and exchange of a large number of image annotations in form of spatial coordinates~\cite{supp222}.
The graphic types used in the ANN objects have been harmonized with those in SRs, and their structure is similar to that of SEG images.
This supplement was recently approved and incorporated into the DICOM standard and is now implemented in \emph{highdicom} as a \texttt{highdicom.ann.MicroscopyBulkSimpleAnnotations} SOP class, reusing the existing building blocks of the library for coded concepts and spatial coordinates.

\subsection*{Abstracting the complexity of the standard without oversimplifying medical imaging use cases}

DICOM is the ubiquitous standard for representation and communication of medical image data and standardizes many aspects of the imaging workflow to enable interoperability in the clinical setting.
However, DICOM is often criticized by the biomedical imaging research community for its elaborateness and alternative data formats have emerged in the research setting that are intended to simplify access to and storage of data by researchers that do not want to cope with intricacies of the standard~\cite{pmid24338090}.
The first step in an image analysis pipeline is thus often the conversion of DICOM objects into an alternative format that is considered more suitable for research use~\cite{pmid26945974}.
While conversion of clinically acquired DICOM objects into another format may work well within the limited scope of a research project, the reverse, i.e., the conversion of a given research output into standard DICOM representation, is generally not possible, since important contextual information is lost along the way~\cite{roberts2021common}.
Many of the attributes of DICOM objects that are regarded superfluous by researchers and are readily removed for ease of use, are crucial for interoperability with clinical systems and for correct representation and interpretation of the data in clinical practice.

We argue that the discussion regarding the establishment and adoption of standards for clinical deployment of ML models and integration of their outputs into clinical workflows should be guided primarily by the requirements of clinical systems and clinicians for interpretability and clinical decision making, rather than current practices within research communities.
The DICOM standard has been evolving over many years through continuous collaboration of an international group of experts and a diverse set of stakeholders based on a considerate and controlled process that takes a variety of use cases as well as legal and regulatory aspects into account.
While the comprehensiveness and inclusiveness of the standard has advantages, it has also resulted in significant complexity and demands an implementation that exposes the useful parts of the standards through a layer of abstraction.
The \emph{highdicom} library strikes a fine balance, by providing an API that hides as many details of the DICOM standard as possible from model developers, while acknowledging that medical imaging is complex and that efforts aiming for DICOM abstraction should involve technical and domain experts to avoid oversimplification with detrimental effects on interoperability and ultimately patient safety.
The result is an API that abstracts the intricate structure of DICOM data sets, but retains full and direct access to all DICOM attributes and stays close to the terminology of the DICOM data models to avoid any ambiguities.

\subsection*{Bridging the gap between model development in research and model deployment in clinical practice}

Researchers, medical device manufacturers, and healthcare providers are generally interested in accelerating the translation of research findings into clinical practice and enable patients to get access to and benefit from diagnostic and therapeutic innovations.
However, the incentives for the different stakeholders who participate in the translation process at different time points from model development to deployment are not necessary well aligned.
Currently, the production deployment of an ML model is generally not a major concern to model developers, who primarily operate in a research environment.
The developer often does not receive a technical specification against which the model should be developed and is unaware of the environment into which the model should ultimately be deployed for clinical validation.
As a consequence, the structure of data outputted by ML models developed in research settings are generally highly customized towards a particular research project and specific use case and lack identifying or descriptive metadata relevant for clinical application (see above).
Further, current ML models store data in a variety of proprietary formats that are incompatible with clinical systems, which generally rely on a DICOM interface for data exchange.
Together, these factors impede the deployment of an ML model and its integration into existing clinical workflows for validation or application.

One opportunity for streamlining this process is to rely on DICOM as a common format and interface for data exchange during both model development and deployment.
In our experiments, we demonstrated that \emph{highdicom} makes feasible a fully DICOM-based workflow in which all files stored on storage devices are in DICOM format with minimal increase in complexity for the developer.
Adapting a model developed in such a workflow for clinical deployment becomes a straightforward task.

A common use for non-DICOM formats is for storage of intermediate results within the input image preprocessing pipeline, such as the results of image registration operations.
A limitation of our proposed DICOM-only workflow is that it assumes that model training and inference pre-processing pipelines operate directly on the source images.
However, we argue that models developed for eventual clinical deployment must have input preprocessing pipelines that are able to operate efficiently from the raw source data and as such having this constraint in place through model development process simplifies deployment.
Furthermore, intermediate results could also be represented in DICOM format (e.g., using the Spatial Registration IOD for image registration results) and future versions of \emph{highdicom} may provide tools to help with the creation and access of intermediate results in DICOM format.

\subsection*{Common platforms, services, and tools will facilitate enterprise medical imaging, interdisciplinary research, and integrated diagnostics}

Standardization of images, image annotations, and model predictions between pathology and radiology opens new avenues for enterprise medical imaging, interdisciplinary quantitative biomedical imaging research, and integrated image-based diagnostics.
Despite unique challenges and use cases for image management in digital pathology and radiology, there are opportunities for streamlining the investment into and use of IT infrastructure and platforms across medical disciplines within the enterprise.
Given that most hospitals already have an existing medical imaging infrastructure based on DICOM, encoding image annotations in DICOM format may lower the barrier for integration of ML systems into clinical workflows.

Relying on the DICOM standard may further promote interdisciplinary biomedical imaging research by, for example, clearing the way for the use of annotations of slide microscopy images in pathology as ground truth for training ML models for analysis of CT images in radiology or vice versa.
Furthermore, leveraging a standard data format and communication interface provides an opportunity to synthesize different imaging modalities and interpret pathology and radiology ML model outputs side-by-side.
In this paper, we demonstrate that \emph{highdicom} facilitates the creation and interpretation of image annotations independent of a specific medical imaging modality, discipline, department, or institution.
We further show that data can be exchanged and stored using DICOM-compatible image management systems, which already exist in hospitals worldwide and are increasingly being adopted by biomedical imaging research initiatives around the world.
For example, the National Cancer Institute's Imaging Data Commons (IDC) in the United States will make large public collections of pathology and radiology images, image annotations, and image analysis results available in DICOM format~\cite{idc}.
The \emph{highdicom} library will allow researchers to leverage these resources and enable them to readily share their results and make them usable by other researchers.
We therefore see the potential for \emph{highdicom} to streamline the development and deployment of ML models across departmental boundaries, accelerate the translation of technological innovations from research into clinical practice, and to assist in the realization of AI in healthcare.

%% file: sections/conclusion.tex
The \emph{highdicom} library abstracts the complexity of the DICOM standard, and exposes medical imaging data to ML model developers via a pythonic interface that ties into the scientific Python ecosystem for machine learning and image processing and allows data scientists to think of imaging data at a high level of abstraction without having to worry about the low-level details and rules of the DICOM standard.
Focusing on the use case of detecting lung tumors in slide microscopy images of surgical tissue section specimens as well as in computed tomography images of the chest, we examined examples for the interpretation of DICOM-encoded image annotations during model training and encoding of model outputs during model inference.
Through a series of experiments, we have demonstrated the utility of the library for the development of ML models and shown that, by relying on the DICOM standard, the library enables interoperability of the developed ML models with commercially available DICOM-compliant information systems and allows for unambiguous interpretation of model outputs in clinical context independent of the specific medical imaging modality or discipline.
By facilitating the use of DICOM throughout the model development and deployment process, \emph{highdicom} has the potential to bridge the gap between research and clinical application and thereby streamline clinical integration and validation of ML models.